\documentclass[twocolumn,showpacs,superscriptaddress,preprintnumbers,amsmath,amssymb,aps,reprint]{revtex4-1}
\usepackage{graphicx}
\usepackage{mathptmx}
\usepackage{verbatim}
\usepackage{graphicx}
\usepackage{amsmath}
\usepackage{amssymb}
\usepackage{amsfonts}
\usepackage{mathrsfs}
\usepackage{amsmath}     
\usepackage{color}
\usepackage{bm}               
\usepackage{amssymb}
\usepackage{xspace} 
\usepackage{hyperref}
\usepackage{multirow}
\usepackage{color}
\usepackage{float}
\usepackage{bbold}
\usepackage{tabu}
\usepackage{textcomp}
\usepackage{multirow}
\usepackage{dcolumn}
\usepackage{float}
\usepackage{appendix} 

\newcolumntype{L}[1]{>{\raggedright\arraybackslash}m{#1}}
\newcolumntype{C}[1]{>{\centering\arraybackslash}m{#1}}
\newcolumntype{R}[1]{>{\raggedleft\arraybackslash}m{#1}}
\newcolumntype{N}{@{}m{0pt}@{}}

\usepackage{listings}
\usepackage{color}
\usepackage[utf8]{inputenc}

\definecolor{dkgreen}{rgb}{0,0.6,0}
\definecolor{gray}{rgb}{0.5,0.5,0.5}
\definecolor{mauve}{rgb}{0.58,0,0.82}

\begin{document}
\title{Atomistic substrate relaxation effects in the band gaps of graphene on hexagonal boron nitride}

\author{Jiaqi An}
\affiliation{Department of Physics, University of Seoul, Seoul 02504, Korea}
\author{Nicolas Leconte}
\altaffiliation{Current affiliation: Catalan Institute of Nanoscience and Nanotechnology (ICN2), CSIC and BIST, Campus UAB, Bellaterra, 08193 Barcelona, Spain}
\author{Srivani Javvaji}
\affiliation{Department of Physics, University of Seoul, Seoul 02504, Korea}
\author{Youngju Park}
\affiliation{Department of Physics, University of Seoul, Seoul 02504, Korea}
\author{Jeil Jung}
\email{jeiljung@uos.ac.kr}
\affiliation{Department of Physics, University of Seoul, Seoul 02504, Korea}

\date{\today}
\begin{abstract}
We assess the impact of atomistic substrate lattice relaxation schemes in the primary band gap at charge neutrality and the secondary valence band gap of graphene on hexagonal boron nitride (G/h-BN) as a function of twist angle. 
For zero twist angle, the primary gap decreases from $\sim 30$~meV in fully relaxed suspended G/h-BN bilayers, to $\sim 9$~meV when the remote h-BN substrate layer is kept rigid, and down to $\sim 3$~meV in completely rigid structures. 
In the presence of relaxations, the primary gap shows a maximum near $\sim 0.6^{\circ}$ coinciding with energetic stabilization due to alignment between the moir\'e pattern and the graphene lattice vectors, while
the secondary valence band gap drops from $\sim 12$~meV down to zero beyond twist angles of $\sim 1^{\circ}$. 
A small but finite primary gap on the order of $\sim 1$~meV, with a mass sign favoring electronic occupation of carbon atop boron, persists across twist angles from $0^{\circ}$ to $30^{\circ}$ for all sliding configurations, and switches sign for twist angles between $30^{\circ}$ and $60^{\circ}$.

\end{abstract}
\maketitle

\section{Introduction}
Research on two-dimensional (2D) materials such as graphene, hexagonal boron nitride (h-BN), and transition metal dichalcogenides (TMDs) has remained highly active over the past decades due to the novel material properties emerging in the ultrathin limit~\cite{Meng2004,Science2008,doi:10.1021/nl302015v,doi:10.1021/nl303583v,Li2014,PhysRevLett.112.176801,doi:10.1126/science.aab3175,https://doi.org/10.1002/adfm.201501972,doi:10.1126/science.aad0201,doi:10.1126/science.aav3548,Leconte2020,Zheng2020}. 
A prototypical van der Waals heterostructure in this class is G/h-BN, whose electronic properties have been extensively studied over the last decade~\cite{doi:10.1126/science.1237240,0a7c5baa122e4f1f9f96b95b0f35eb3e,Wallbank:2013ep}.
One of the hallmark features of G/h-BN heterostructures is the emergence of electronic band gaps near the Dirac point and at higher energies, arising from the moiré potential induced by the lattice mismatch between graphene and h-BN. These moiré-induced effects have been confirmed in multiple experiments~\cite{woods, Amet:2013gw, Hunt:2013ef,Wang:2016ii,Yankowitz:2014bv,doi:10.1126/science.1237240,0a7c5baa122e4f1f9f96b95b0f35eb3e,Wallbank:2013ep,Chen2014,10.1021/acs.nanolett.8b03423}, which observed both the first (primary) Dirac point (FDP) and the secondary Dirac point (SDP) features.
Experimental and theoretical studies have shown that these band gaps depend sensitively on the twist angle between graphene and h-BN, with reported behaviors ranging from linear~\cite{Ni2019,PhysRevLett.111.266801} to nonlinear~\cite{RibeiroPalau2018,PhysRevB.90.075428} reductions as the twist angle increases. 
For aligned G/h-BN, the primary and secondary gaps have been measured to be approximately $14$~meV via tunneling spectroscopy~\cite{10.1021/acs.nanolett.8b03423}, in reasonable agreement with $\sim 8$~meV obtained from a DFT-inspired continuum model~\cite{jung2013_PRB}, which increases by about a factor of two upon inclusion of Hartree-Fock interactions.
Other theoretical works incorporating atomistic lattice relaxation in real space have reported larger primary gap values, in the range of $20 \sim 35$~meV, depending on the details of the structural relaxation scheme and the electronic structure model~\cite{PhysRevB.100.195413,Long2022}. 
This apparent discrepancy between continuum model predictions, experimental results, and atomistic simulations motivates our investigation into the quantitative role of substrate relaxation schemes in determining the electronic gaps of G/h-BN.

We mainly focus on the evolution of the primary and secondary band gaps under different lattice relaxation schemes of the substrate h-BN which has been neglected in past atomistic relaxation studies~\cite{PhysRevB.100.195413,Long2022}. 
By systematically varying the relaxation constraints on the substrate atoms, we assess how the treatment of lattice reconstruction influences the gap behavior across twist angles.
We find that when the remote h-NB substrate layers are held rigid, the total energy exhibits a shallow minimum at a finite twist angle of $\sim 0.6^{\circ}$. Concurrently, the primary band gap shows a weak local enhancement at this angle. As the twist angle increases, the primary gap gradually decreases, reaching values on the order of $\sim 1$~meV for large angles without fully closing. The actual values at larger angle are sensitive on whether the system is even or odd~\cite{PhysRevB.81.161405}, \textit{i.e.} the rotation center is chosen at the AA or AB high-symmetry stacking.
This local energy minimum and corresponding gap enhancement coincide with a regime of increased lattice reconstruction, characterized by an expansion of the BA-stacking region and a locally enhanced rotation angle in the AA-stacking zone, relative to the perfectly aligned case. While similar reconstruction patterns are observed in suspended (free-standing) G/h-BN systems, they do not lead to a comparable energy minimum.
The origin of this energy minimum can be traced to a commensuration effect: in lattice-mismatched systems, the moiré lattice vectors align with those of the graphene layer at a small but finite twist angle. For graphene on h-BN, this alignment occurs near $0.6^\circ$, resulting in enhanced structural stability.~\cite{jharapla2025geometriccontrolmoiretwist} 

The paper is organized as follows. In Section~\ref{sec:System}, we describe our method for constructing commensurate supercells for G/h-BN heterostructures using a set of four integer indices. 
Section~\ref{TBSect} introduces our tight-binding (TB) model for the electronic structure of G/h-BN, {as well as the truncated atomic plane wave(TAPW) method  used to compute the average mass term; the parameters for atomic structure relaxation via the LAMMPS package~\cite{10.1371/journal.pcbi.1004410} are also detailed.} 
In Section~\ref{sec:Result}, we present our main results for the twist angle dependence of the primary and secondary band gaps, highlighting how different relaxation schemes influence their magnitude and evolution.

\section{System}
\label{sec:System}

\begin{figure}[tb]
    \includegraphics[width=1.0\columnwidth]{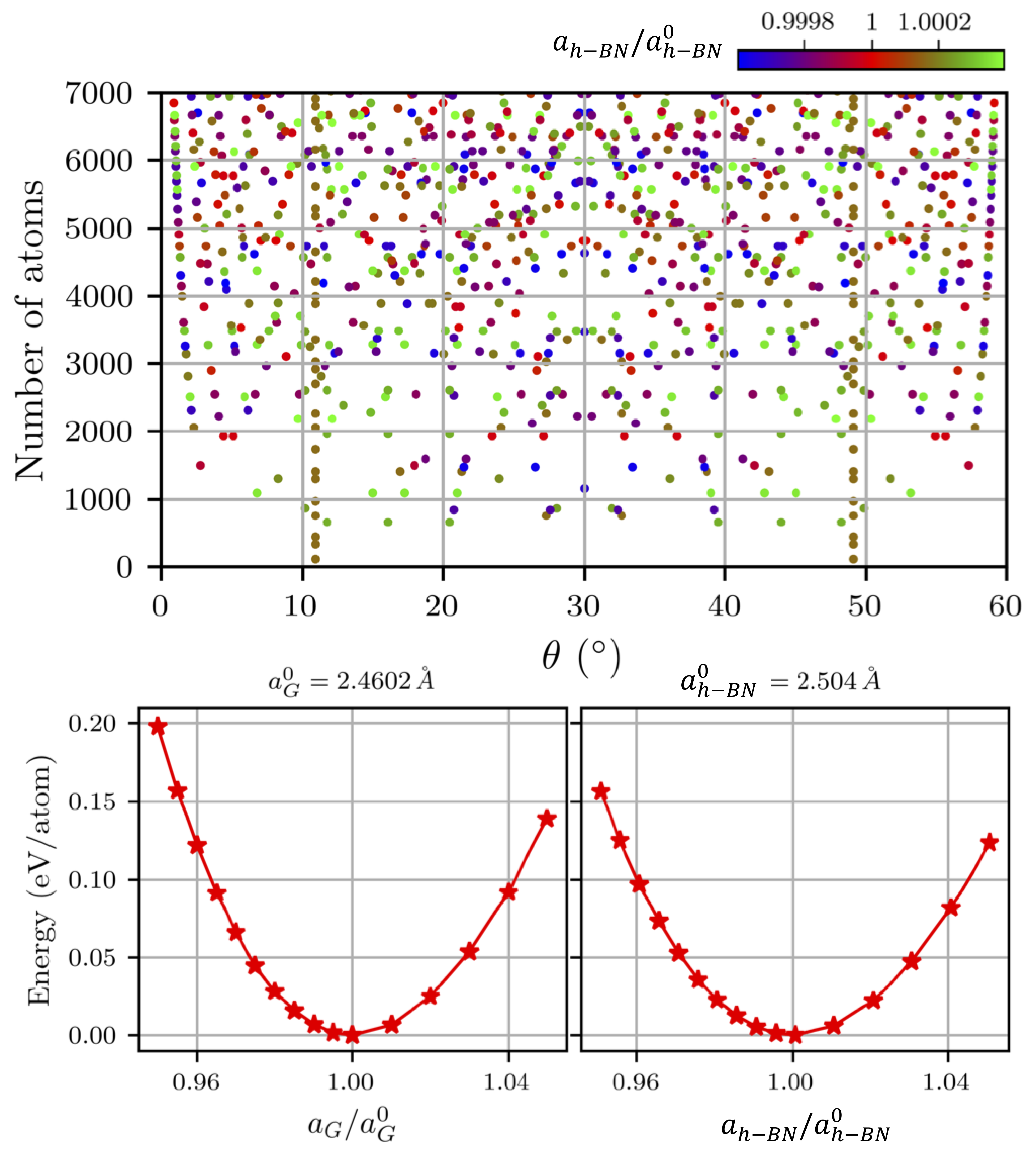}
\caption{(Color online) Number of atoms in the commensurate supercells constructed using the four indices defined in Eq.~\ref{twistangle}. A lattice tolerance of $\pm 0.01$~\AA\ is allowed with respect to the equilibrium h-BN lattice constant $a_{h-BN}^0 = 2.504$~\AA, increasing the density of accessible twist angles via smaller commensurate cells. The graphene lattice constant is held fixed at $a_G^0 = 2.4602$~\AA. The bottom panels show the elastic energy correction per atom, calculated using Eq.~(\ref{equlibriumDistance}), when the graphene or h-BN lattice constants are varied from their equilibrium values using the same force fields employed in the energy minimization. Total energies are shifted by $-7.394$~eV/atom for graphene and $-6.690$~eV/atom for h-BN to set the energy minimum to zero.}
    \label{HermannGBN}
\end{figure}

\begin{figure}[tbhp]
    \includegraphics[width=0.49\textwidth]{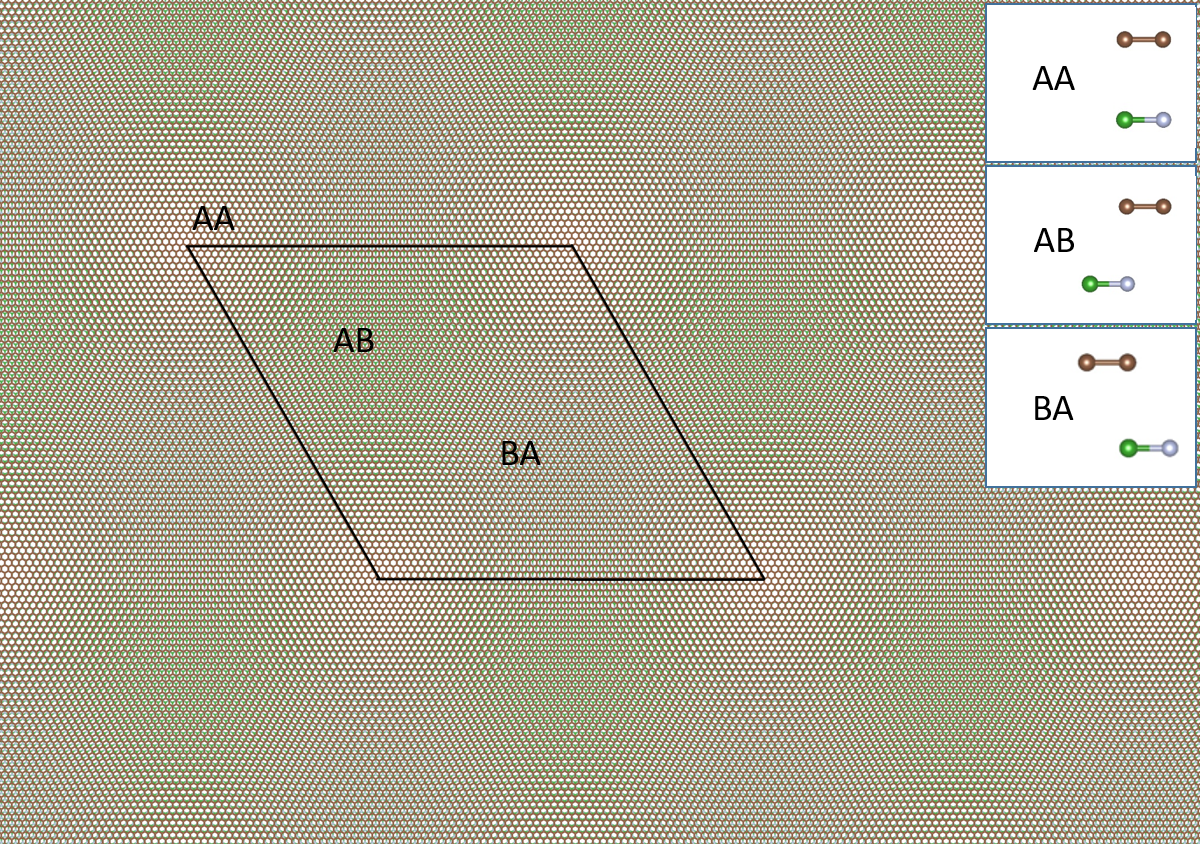}
    \caption{(color online) Top view of the aligned G/h-BN moir\'e pattern generated with indices $i = 55, j = 0, i^\prime = 54, j^\prime = 0$. In the AA stacking region, the carbon atoms are on top of boron and nitrogen atoms, respectively. In the AB region, the carbon atom is on top of the nitrogen atom and the center of the graphene hexagon is on top of the boron atom. In the BA region, the role of B and N is reversed with respect to AB. The moir\'e length is $L_M = 135.311$~\AA\ for zero angle and decreases for increasing twist angles following Eq.~(\ref{eq:MoireLength2}).}
    \label{commensuration}
\end{figure}

We construct commensurate moiré supercells for G/h-BN heterostructures using four integers $p$, $q$, $p^\prime$, and $q^\prime$, following an approach similar to that used for tBG~\cite{https://doi.org/10.48550/arxiv.1910.12805}, based on the methodology outlined in Ref.~[\onlinecite{Hermann2012}]. 
The lattice vectors of the commensurate supercell, ${\bm r}_{1}$ and ${\bm r}_{2}$, are related to the lattice vectors of the bottom layer, ${\bm a}_{1}$ and ${\bm a}_{2}$, and the top layer, ${\bm a}^\prime_{1}$ and ${\bm a}^\prime_{2}$, via

\begin{equation}
\begin{pmatrix}
{\bm r}_{1} \\
{\bm r}_{2}
\end{pmatrix}
= {\bm M}^{\prime} \cdot
\begin{pmatrix}
{\bm a}_{1} \\
{\bm a}_{2}
\end{pmatrix}
= {\bm M} \cdot
\begin{pmatrix}
{\bm a}^{\prime}_{1} \\
{\bm a}^{\prime}_{2} 
\end{pmatrix},
\end{equation}
where the transformation matrices ${\bm M}$ and ${\bm M}^\prime$ are given by
\begin{equation}
{\bm M} =
\begin{pmatrix}
p & q \\
- q & p + q
\end{pmatrix}, 
\quad
{\bm M}^{\prime} =
\begin{pmatrix}
p^\prime & q^\prime \\
- q^\prime & p^\prime + q^\prime
\end{pmatrix}.
\end{equation}
Let $a_G$ and $a_{h-BN}$ denote the lattice constants of graphene and h-BN, respectively. The lattice mismatch ratio $\alpha$ between the two layers is then expressed in terms of the integer indices as
\begin{equation}
\alpha = \frac{a_G}{a_{h-BN}} =\sqrt{\frac{p^{\prime 2}+q^{\prime 2}+p^\prime q^\prime}{p^2+q^2+pq}},
\label{latticemismatch}
\end{equation}
and the twist angle $\theta$ is given by
\begin{equation}
\cos(\theta) = \frac{1}{2 \alpha g}\left( 2 p^\prime p + 2 q^\prime q + p^\prime q + q^\prime p \right),
\label{twistangle}
\end{equation}
where $g = p^2 + q^2 + pq$.
In Fig.~\ref{HermannGBN}, we illustrate the set of commensurate cells accessible up to $60^\circ$ twist angles, allowing for a tolerance of $0.001$~\AA{} on the h-BN lattice constant $a_{h-BN}$. This tolerance increases the density of accessible twist angles using smaller commensurate cells, while the resulting internal strain remains negligible for the observables reported in this work. However, for total-energy comparisons between angles, higher precision is required to minimize lattice mismatch noise. We thus use larger simulation cells for this purpose, as illustrated in the Appendix.
In the lower panel of Fig.~\ref{HermannGBN}, we show the elastic energy penalty incurred when the lattice constants of graphene and h-BN are varied from their respective equilibrium values. These penalties are fitted to parabolic functions of the form
\begin{equation}
\begin{split}
E_{G} &= 11.44 \, (a_{G} - a^{0}_{G})^2, \\
E_{h-BN} &= 9.11 \, (a_{h-BN} - a^{0}_{h-BN})^2,
\end{split}
\label{equlibriumDistance}
\end{equation}
where $a^{0}_{G} = 2.4602$~\AA\ and $a^{0}_{h-BN} = 2.504$~\AA, and energies are expressed in eV/atom. The prefactors are given in units of $\rm{eV}/\AA^2$. Under this convention, a relative change of $0.03\%$ in the lattice constant results in an elastic energy penalty of approximately $0.006$~meV/atom for graphene and $0.005$~meV/atom for h-BN.
As a representative case, we consider the aligned G/h-BN configuration defined by the integer indices $p = 55$, $q = 0$, $p^{\prime} = 54$, and $q^{\prime} = 0$, which yields the commensurate supercell illustrated in Fig.~\ref{commensuration}, with a lattice mismatch of approximately $1.8\%$. To minimize internal strain while preserving a manageable simulation cell size for our exact diagonalization calculations, we use $a_{G} = 2.4602$~\AA\ and $a_{h-BN} = 2.50576$~\AA, values chosen to match the equilibrium lattice constants obtained from the force fields used to minimize the energy of the system, as introduced in Section~\ref{MDSection}.
Table~\ref{Commensurate} summarizes the set of parameters used to generate the commensurate supercells for the twisted G/h-BN systems considered in this work.
\begin{table}[]
 \resizebox{1.0\columnwidth}{!}{%
\begin{tabular}{|c|c|c|c|c|c|c|}
\hline
$\theta$ ($^{\circ}$) & ($p$ $q$ $p'$ $q'$) & $a_{h-BN}/a_{h-BN}^0-1$ (\%) & \# atoms & $L_M$($\AA$) & $L_C$($\AA$)  & $\Delta$ E (meV/atom) \\\hline
0.0   & 54 0 55 0   & 0.0703  & 11882       & 135.311   & 135.311 & 5.25E-3   \\ \hline
0.1   & 51 6 52 6   & 0.0354  & 11990       & 135.914   & 135.914 & 1.08E-3  \\ \hline
0.2   & 46 12 47 12 & 0.0138  & 11458       & 132.874   & 132.874 & 4.61E-3  \\ \hline
0.3   & 42 17 43 17 & 0.0284  & 11272       & 131.776   & 131.776 & 1.66E-4  \\ \hline        
0.396 & 37 21 38 21 & 0.0054  & 10540       & 127.433   & 127.433 & 1.70E-3  \\ \hline
0.5   & 32 24 33 24 & 0.0402  & 9650        & 121.948   & 121.948 & 9.22E-3  \\ \hline              
0.558 & 29 26 30 26 & 0.0327  & 9254        & 119.415   & 119.415 & 6.10E-3   \\ \hline
0.769 & 30 20 30 21 & 0.0297  & 7742        & 109.223   & 109.223 & 5.06E-3   \\ \hline
0.915 & 31 15 31 16 & 0.0386 & 6728        & 101.824   & 101.824 & 8.54E-3   \\ \hline
1.381 & 39 26 41 25 & 0.0299  & 13084       & 81.978    & 141.990 & 5.11E-3   \\ \hline
1.530 & 34 27 36 26 & 0.0342  & 11422       & 76.596    & 132.668 & 6.68E-3   \\ \hline
2.060 & 47 4 49 2   & 0.0263  & 9832        & 61.542    & 123.084 & 3.97E-3   \\ \hline
2.558 & 48 11 47 14 & 0.0241  & 12032       & 51.463    & 136.158 & 3.31E-3   \\ \hline
3.071 & 43 6 42 9   & 0.0279  & 8732        & 43.842    & 115.995 & 4.46E-3   \\ \hline
4.004 & 49 39 55 34 & 0.0312  & 23768       & 34.372    & 191.372 & 5.15E-3   \\ \hline
4.996 & 48 57 57 50 & 0.0300  & 33776       & 27.871    & 228.136 & 5.57E-3   \\ \hline
10.02 & 27 35 37 26 & 0.0242  & 11812       & 14.142    & 134.908 & 3.36E-3   \\ \hline
15.08 & 26 15 17 25 & 0.0207  & 5260        & 9.437     & 90.025 & 2.44E-3   \\ \hline
20.05 & 20 12 28 1  & 0.0115  & 3194        & 7.122     & 70.148 & 7.56E-4   \\ \hline
24.99  & 9 17 19 7   & 0.0718 & 2132        & 5.733    & 57.328 & 2.95E-2   \\ \hline
30.00 & 17 0 10 10  & 0.0632  & 1178        & 4.794     & 42.612 & 2.28E-2   \\ \hline
\end{tabular}
}
\caption{Commensurate twist angles for G/h-BN along with their corresponding integer indices defined in Eq.~(\ref{twistangle}). The table also includes the adjusted lattice constant $a_{h-BN}$, the number of atoms in the resulting supercell, and the corresponding moiré length $L_M$ computed from Eq.~(\ref{eq:MoireLength2}).}
\label{Commensurate}
\end{table}
The relationship between the moiré length $L_M$ and the twist angle $\theta$ is given by~\cite{Shin2021, Leconte2020}
\begin{equation}
    L_{M} = \frac{a_{G}}{\sqrt{\alpha^2 - 2\alpha \cos(\theta) + 1}},
    \label{eq:MoireLength2}
\end{equation}
where $\alpha = a_G / a_{h-BN}$ is the lattice mismatch ratio. The corresponding $L_M$ values for the commensurate structures studied are listed in Table~\ref{Commensurate}. We note that the even/odd counterparts~\cite{PhysRevB.81.161405} of the systems with angles $\theta<30^\circ$, corresponding to the systems $\theta = 60^\circ - \theta > 30^\circ$, can be generated using the same set of indices by interchanging the roles of B and N atoms.

\section{Tight-binding model and MD simulations}
\label{TBSect}

The tight-binding electronic structure model for G/h-BN we propose here separates the interlayer and intralayer contributions 
\begin{equation}
    {H} = {H}_\text{intra} + {H}_\text{inter}
\label{eq:Hamiltonian}
\end{equation}
where the latter uses a refined two-center (TC) model based on the inter-atomic distance-vector ${\bm r}_{kl}$ as introduced in Ref.~[\onlinecite{https://doi.org/10.48550/arxiv.1910.12805}] under the Slater-Koster approximation~\cite{laissardire2012, nam2017, Moon2014} and where the former utilizes accurate short range F2G2 models based on DFT Wannierization of bands where their moir\'e-dependence is captured by the displacement vector ${\bm d}({\bm r_{kl}}) = (\alpha \mathcal{R}(\theta)-1){\bm r_{kl}}$ where $\alpha = a_G/a_{h-BN}$ and $\mathcal{R}(\theta)$ is the rotation operator.
The $k$ and $l$ indices in ${\bm r}_{kl}$ are the atom indices for any atom pair considered in the long-range TC interlayer model.
The vector $\bm{d}(\bm{r_{kl}})$, where we omit the $\bm{r}$-lattice-dependence from hereon to simplify notation, is obtained by finding the vector that connects the atom at $\bm{r}$ with its closest interlayer neighbor that is in the same sub-lattice.
To capture the fact that we replace the conventional intralayer TC terms with more accurate short-range F2G2 terms, we will refer to this model as a hybrid TC (HTC) model.

\subsection{Intra-layer contributions}
By using  intra-layer F2G2 terms in our HTC model, we improve on the conventional intra-layer TC model term, where moir\'e effects due to the contacing layer are introduced based on \textit{ab initio} calculation as presented in Ref.~[\onlinecite{Jung:2014hj}] where we can systematically control the range of the finite hopping terms.
We summarize these intralayer terms that are valid close to the ${\bm K}$ high-symmetry k-point as
\begin{equation}
    H_\text{intra}({\bm d}) = H_{ii}({\bm d}) + H_{ij}({\bm d})
    \label{intraEq}
\end{equation}
where the inter-sublattice terms are given as
\begin{equation}
    H_{ij}({\bm d}) = \sum_n t_{{ij}_n } ({\bm d}) f_n
    \label{interEq}
\end{equation}
and intra-sublattice terms
\begin{equation}
    H_{ii}({\bm d}) = \sum_n t_{{ii}_n} ({\bm d}) g_n.
    \label{OnsiteEq}
\end{equation}
where $i$ and $j$ refer to intra-layer atoms.
For each interaction type, we can express the intralayer Hamiltonian as a sum of the product between the hopping strength $t_{{ij}_n}$ 
and the corresponding structure factors $f_n$ or $g_n$ that are defined in Ref.~[\onlinecite{jung2013_PRB}].
We summarize the $t_{{ij}_n}$ factors in Table~\ref{table:elasticCoeff} by their average value obtained from calculations performed for the three main AA, AB and BA stacking configurations. We note that here we use bilayer G/h-BN DFT data to parametrize the F2G2 terms as it best reproduces the band structures at the high symmetry stacking points, but single-layer-parametrized F2G2 models can also be used to accurately simulate the low energy range, as illustrated in Appendix~\ref{AppendixF2G2}. These average values are used for all interactions except for the onsite energies and the first-nearest neighbor hopping terms, where we additionally include the moir\'e-dependence through ${\bm d}$.
\begin{table}[tbhp]
\begin{center}
\begin{tabular}{|c|c|c|c|c|c|c|c|}
\hline
$t_{{ij}_n}$         & $C_1-C_1$ & $C_2-C_2$  & B-B       & N-N       && $C_1-C_2$ & B-N       \\ \hline
$\ g_0\ $    & -0.4017   & -0.4027    & 2.3143    & -1.7966   &&           &           \\ \hline
$\ g_1\ $    & 0.24498   & 0.24523    & 0.081055  & 0.24562   &$\ f_1\ $& -3.0307   & -2.8928   \\ \hline
$\ g_2\ $    & 0.06618   & 0.06624    & 0.065654  & 0.04892   &$\ f_2\ $& -0.19334  & -0.15399  \\ \hline
\end{tabular}
\caption{\label{table:elasticCoeff} Average hopping terms $t_{{ij}_n}$ corresponding to the intrasublattice terms with structure factors $g_n$ and intersublatice interactions with structure factors $f_n$ as defined in Ref.~[\onlinecite{jung2013_PRB}]. $g_0$ refers to the onsite energy and $f_1$ links first-nearest neighbor atoms. The average is taken over the values obtained for the three different AA, AB and BA stacking configurations where they are nearly the same except for the $g_0$ terms, as illustrated in Table~\ref{Table:OnsiteABC}.
}
\end{center}
\end{table}
\begin{table}[tbhp]
\begin{center}
\resizebox{0.7\columnwidth}{!}{%
\begin{tabular}{|c|c|c|c|c|}
\hline
       &    A      &     B      &        C          &  C$_{0ii}$   \\ \hline
$H_{{\rm C}_1{\rm C}_1}$ & -0.39969  &  -0.37776  &  -0.4277  & -0.4017    \\ \hline
$H_{{\rm C}_2{\rm C}_2}$ & -0.42011  &  -0.36739  &  -0.42048 & -0.4027   \\ \hline
$H_{BB}                $ & 2.3222      & 2.2731  & 2.3476   &  2.3143     \\ \hline
$H_{NN}                $ & -1.7795    & -1.8324   & -1.7778 &  -1.7966   \\ \hline
\end{tabular}}
\caption{\label{Table:OnsiteABC} $H_\text{onsite}$ from Eq.~(\ref{eq:Onsite}) at different stacking configurations AA, AB and BA represented here by A, B and C labels. The average, equal to $C_{0ii}$ in Eq.~(\ref{eq:Onsite}), matches the value from Table~\ref{table:elasticCoeff} and corresponds to the values used in Fig.~\ref{HABandOnsite} to center the colormaps around zero.}
\end{center}
\end{table}
Indeed, on the one hand, for the onsite energies that correspond to the $g_0=1$ structure factor, we use the first harmonic fitting approximation to obtain the intralayer moir\'e induced ${\bm d}$-dependence from the high-symmetry K-point-values calculated for different select highly-symmetric local stacking configurations, leading to
\begin{equation}
H_{ii}({\bm d}) = C_{0ii}+2C_{1ii}\operatorname{Re}\left[f({\bm d})\exp(i\phi_{ii})\right], \\
\label{eq:Onsite}
\end{equation}
with $ii = C_1C_1$, $C_2C_2$, $BB$ or $NN$, and
\begin{equation}
f({\bm d}) = \exp(-iG_1 d_y)+2\exp\left(i\frac{G_1 d_y}{2}\right)\cos\left(\frac{\sqrt{3}}{2}G_1 d_x\right)
\end{equation}
where $d_x$ and $d_y$ are the vector-components of ${\bm d}$ normalized to either $a_G$ or $a_{h-BN}$ depending on which layer the atom is located in and where $G_1 = 4\pi/\sqrt{3}$. The three parameters from Eq.~(\ref{eq:Onsite}) are summarized in Table~\ref{Table:OnsiteHarmonics}.
\begin{table}[tbhp]
\begin{center}
\resizebox{0.7\columnwidth}{!}{
\begin{tabular}{|c|c|c|c|}
\hline
       &    C$_{0ii}$ (eV)     &     C$_{1ii}$ (meV)     &        $\phi_{ii}$ ($^{\circ}$)   \\ \hline
$H_{{\rm C}_1{\rm C}_1}$ & -0.4017  &  4.8173  &  -85.9793      \\ \hline
$H_{{\rm C}_2{\rm C}_2}$ & -0.4027  &  -5.8784  &  60.3470    \\ \hline
$H_{BB}                $ & 2.3143      & 7.2887  & 79.5926        \\ \hline
$H_{NN}                $ & -1.7966    & 5.9745   & 61.5690    \\ \hline
\end{tabular}}
\caption{\label{Table:OnsiteHarmonics} The coefficients for $H_{ii}({\bm d})$ $g_0$ terms and the expression for this term is Eq.~(\ref{eq:Onsite}). The average, equal to $C_{0ii}$ in Eq.~(\ref{eq:Onsite}), matches the value from Table~\ref{table:elasticCoeff} and corresponds to the values used in Fig.~\ref{HABandOnsite} to center the colormaps around zero.}
\end{center}
\end{table}
We also provide a re-parametrization of these parameters that vary with interlayer distance as
\begin{equation}
 X = a c_\perp^2 + b c_\perp + c  
  \label{quadEq}
\end{equation}
where the fitting parameters $a$, $b$ and $c$ are given in Table~\ref{Table:quadratic} and where $c_\perp$ in our real-space calculations is taken as the vertical distance between the atom under consideration and the plane formed by the atoms in the neighboring layer.

 \begin{table}[th]
 \begin{center}
 \begin{tabular}{|ll|l|l|l|}
\hline
                                      &   & a      & b      & c       \\ \hline
\multicolumn{1}{|l|}{}                & $C_{0ii}$ & -0.085 & 0.651  & -1.622  \\ \cline{2-5} 
\multicolumn{1}{|l|}{$H_{{\rm C}_1{\rm C}_1}$} & $C_{1ii}$ & -12.823 & -97.829  & 188.283  \\ \cline{2-5} 
\multicolumn{1}{|l|}{}                & $\phi_{ii}$ & 64.528 & -398.316  & 583.564  \\ \hline
\multicolumn{1}{|l|}{}                & $C_{0ii}$ & -0.080 & 0.617  & -1.571  \\ \cline{2-5} 
\multicolumn{1}{|l|}{$H_{{\rm C}_2{\rm C}_2}$} & $C_{1ii}$ & 18.394 & -138.487  & 262.558  \\ \cline{2-5} 
\multicolumn{1}{|l|}{}                & $\phi_{ii}$ & 26.432  & -175.793 & 291.727   \\ \hline
\multicolumn{1}{|l|}{}                & $C_{0ii}$ & -0.315 & 2.304 & -1.842 \\ \cline{2-5} 
\multicolumn{1}{|l|}{$H_{BB}$}           & $C_{1ii}$ & -5.067 & 65.747  & -169.785 \\ \cline{2-5} 
\multicolumn{1}{|l|}{}                & $\phi_{ii}$ & -70.528 & 473.619 & -812.156 \\ \hline
\multicolumn{1}{|l|}{}                & $C_{0ii}$ & -1.548 & 10.309  & -18.964  \\ \cline{2-5} 
\multicolumn{1}{|l|}{$H_{NN}$}           & $C_{1ii}$ & -75.783 & 540.730  & -965.290   \\ \cline{2-5} 
\multicolumn{1}{|l|}{}                & $\phi_{ii}$ & 186.49 & -1094.71  & 1579.80  \\ \hline
\multicolumn{1}{|l|}{\multirow{2}{*}{$H_{{\rm C}_1{\rm C}_2}$}}   & $C_{ij}$   & 21.158 & -152.445  & 2256.12   \\ \cline{2-5} 
\multicolumn{1}{|l|}{}                         & $\varphi_{ij}$ & 0.688 & -6.188  & 265.910  \\ \hline
\multicolumn{1}{|l|}{\multirow{2}{*}{$H_{BN}$}} & $C_{ij}$  & 9.112 & -66.125  & 3616.11   \\ \cline{2-5} 
\multicolumn{1}{|l|}{}                         & $\varphi_{ij}$ & 0.974 & -7.907  & 151.089  \\ \hline
\end{tabular}
 \caption{Fitting parameters entering Eq.~(\ref{quadEq}) to take into account the interlayer distance-dependence on the parameters from Table~\ref{Table:OnsiteHarmonics}. The suitable range for the fitting function is [3.1,3.5]. $C_{0ii}$ is in eV unit. $C_{ii}$ and $C_{ij}$ is in meV unit. $\phi_{ij}$ and $\phi_{ii}$ is in $^\circ$ unit.}
 \label{Table:quadratic}
 \end{center}
 \end{table}
\begin{figure}[tbhp]
    \includegraphics[width=1.0\columnwidth]{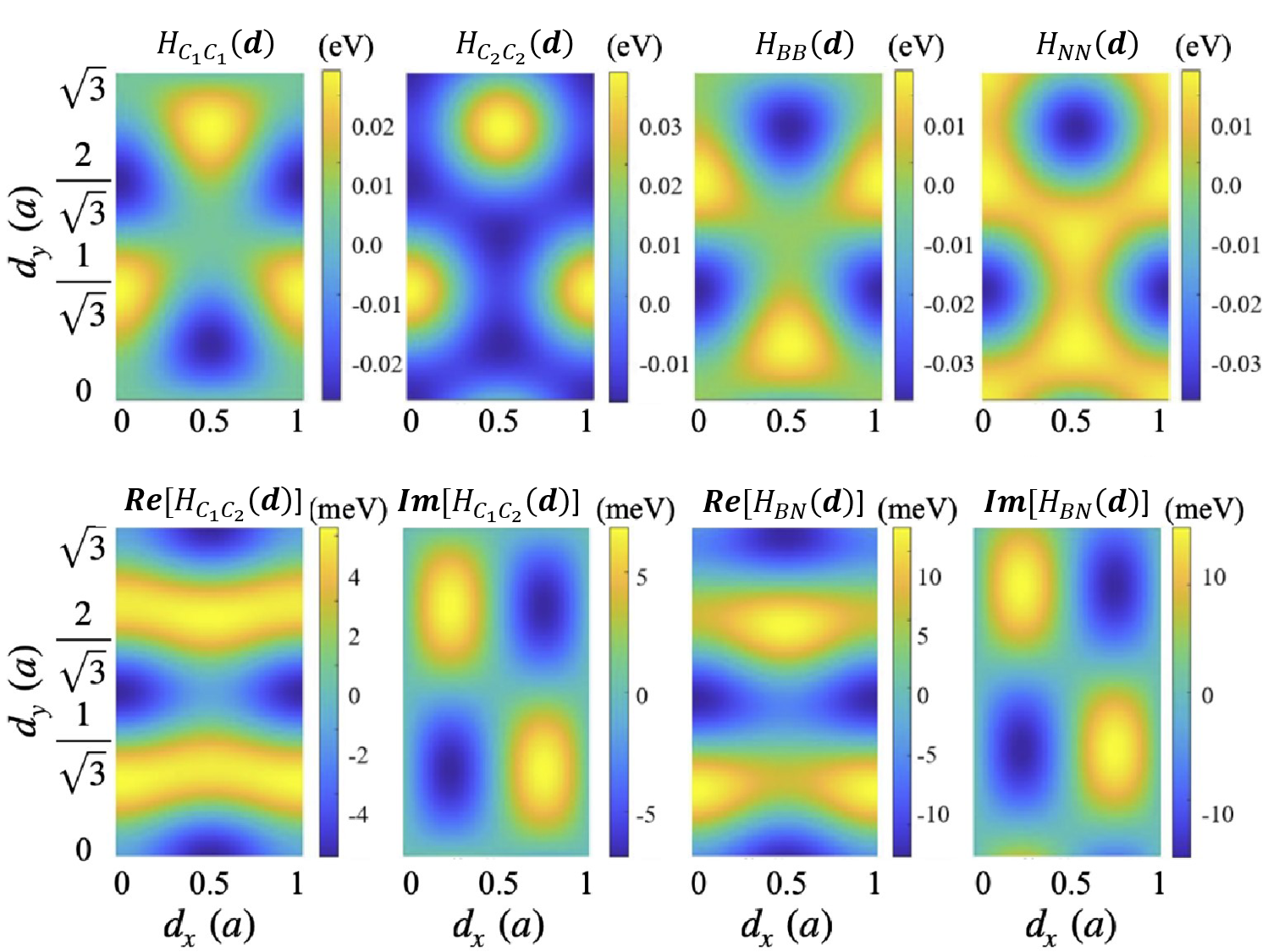}
    \caption{(color online) Matrix elements $H_{ii}({\bm d}({\bm r}_{<kl>}))$ (top panels) and $H_{ij}({\bm d}({\bm r}_{<kl>}))$ (bottom panels) from Eq.~(\ref{eq:Onsite}) and Eq.~(\ref{eq:HAB}) respectively. $C_1$ and $C_2$ are the two sublattices from the graphene layer and  $B$ and $N$ are the atoms from the h-BN layer. We illustrate both the real and imaginary parts of the Hamiltonian. The site energies are plotted relative to their spatial averages given in Table~\ref{Table:OnsiteABC} shifted to zero.}
    \label{HABandOnsite}
\end{figure}
On the other hand, for the first-nearest neighbor hopping terms we introduce the intralayer moir\'e induced ${\bm d}$ -dependence by taking into account a so-called virtual strain correction which captures how the first nearest-neighbor hopping terms are affected by the local stacking configuration with the contacting layer through their ${\bm d}$-dependent off-diagonal $H_{ij}$ terms at the K-point~\cite{jung2014}
\begin{equation}
 \begin{aligned}
 H_{ij}({\bm d}) = 
& 2C_{ij} \cos\left(\frac{\sqrt{3}G_1}{2}d_x\right) \cos\left(\frac{G_1}{2}d_y-\varphi_{ij}\right) \\
& -2C_{ij} \cos\left(G_1 d_y+\varphi_{ij}\right) \\
& -i2\sqrt{3}C_{ij} \sin\left(\frac{\sqrt{3}G_1}{2}d_x\right) \cos\left(\frac{G_1}{2}d_y-\varphi_{ij}\right) \\
\end{aligned}
 \label{eq:HAB}
\end{equation}
where, when assuming a fixed interlayer-distance of $3.35$\AA\ during the DFT paremetrization, $C_{ij} = 1.987$ meV and $\varphi_{ij} = 3.5^\circ$ for the graphene layer ($ij$ corresponding to $C_1-C_2$ interactions) and $C_{ij} = 4.418$ meV and $\varphi_{ij} = 26.10^\circ$ for the h-BN layer ($ij$ corresponding to $B-N$ interactions). Similarly to the onsite energies, a $c_\perp$-dependent parametrization is provided using the fitting fuction from Eq.~(\ref{quadEq}) with the fitting parameters given in Table~\ref{Table:quadratic}. 
We then map this off-diagonal term onto the first nearest-neighbor TB hopping term corresponding to $f_1$ in Table~\ref{table:elasticCoeff} as a $\delta_l$ correction following
\begin{equation}
    t_{{ij}_1} \rightarrow  t_{{ij}_1} + \delta_l 
\end{equation}
where the three nearest neighbors are differentiated in a (counter-)clockwise manner as $l=1$, $2$ and $3$ for (A)B sublattices, respectively~\cite{Leconte2016}
\begin{equation}
\delta_1=2A^{\prime}/3, \delta_2=-(A^{\prime}+\sqrt{3}B^{\prime})/3, \delta_3=(-A^{\prime}+\sqrt{3}B^{\prime})/3
\end{equation}
where $A^{\prime}=\operatorname{Re}(H_{ij}), B^{\prime}=\operatorname{Im}(H_{ij})$.
Finally, to account for (i) the renormalization of the $t_{{ij}_n}$ hopping strengths and the virtual strain correction $\delta_l$ due to the relaxation effects that change the interatomic distances and (ii) the fact that the DFT calculations are performed for a fixed matching lattice constant of $2.439$ \AA\ for 4-atom commensurate cell calculations, we introduce a correction due to global strain as~\cite{Pereira:2009iz} 
\begin{equation}
    t_{{ij}_n} \rightarrow t_{{ij}_n} \exp \left[-3.37\left(\frac{r_{ij}-r_{{0,ij}_n}}{r_{{0,ij}_n}}\right)\right]
\label{t_realStrain}
\end{equation}
and
\begin{equation}
    \delta_l \rightarrow \delta_l \exp \left[-3.37\left(\frac{r_{ij}-r_{{0,ij}_n}}{r_{{0,ij}_n}}\right)\right]
\label{t_realStrain2}
\end{equation}
where $r_{{0,ij}_n}$ are the corresponding unrelaxed inter-atomic distances between intralayer atoms $i$ and $j$ within a triangular lattice with a lattice constant of $2.439$ \AA.
Finally, we note that the DFT Wannier calculations that give us the matrix elements from Eq.~(\ref{interEq}) and Eq.~(\ref{OnsiteEq}) are calculated using on a Monkhorst k-point grid of $42\times42\times1$ with an energy cutoff of $60$~Ry,  
and the associated maps obtained through the harmonic approximation are given in Fig.~\ref{HABandOnsite}. We illustrate in the Appendix~\ref{AppendixA} that this $c_\perp$-dependence of the diagonal and off-diagonal terms has a negligible impact for this system and is thus omitted in the calculations presented in the paper. 

\subsection{Inter-layer contributions}

\begin{figure}[tbhp]
\centering
\includegraphics[width=1.0\columnwidth]{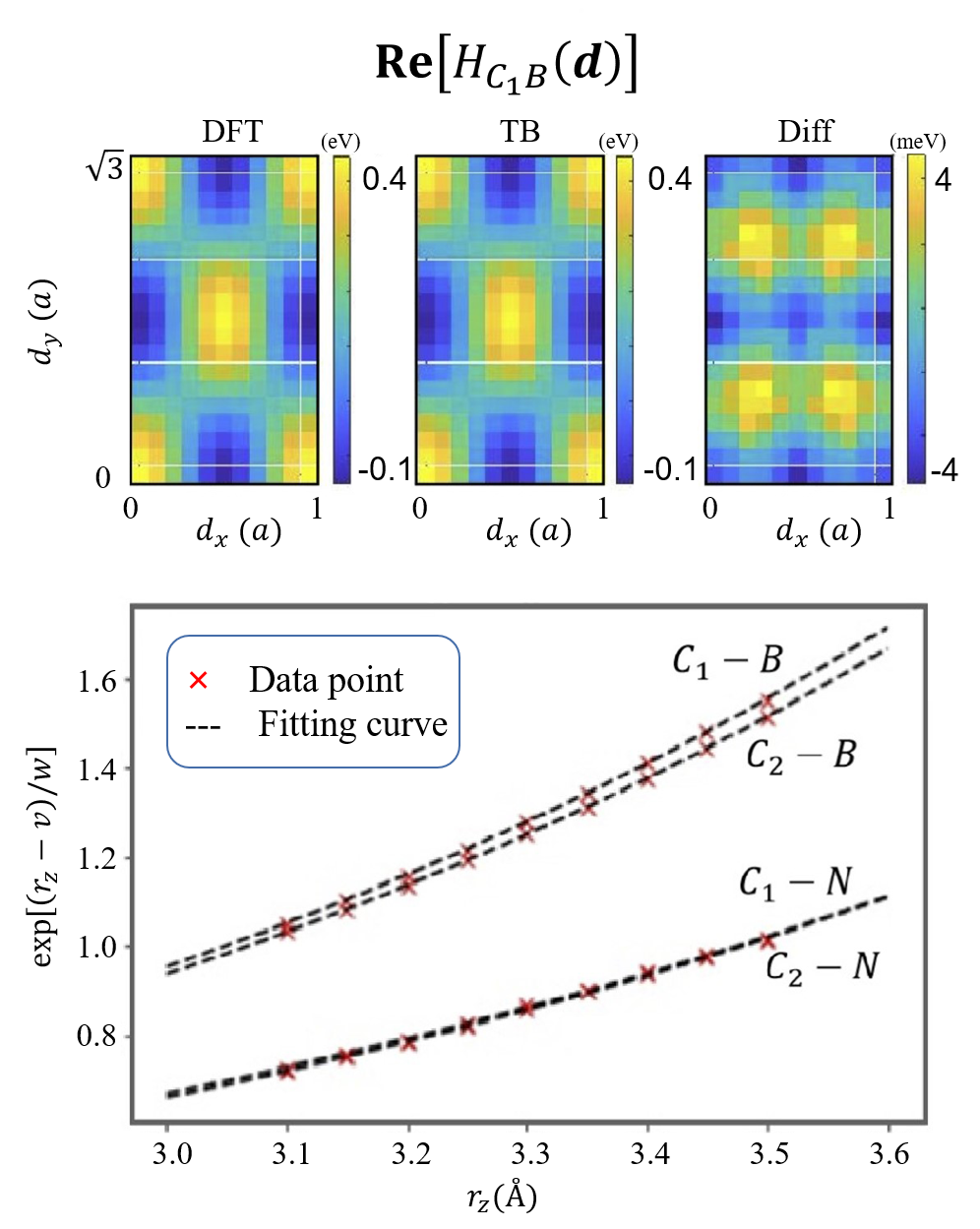}
\caption{\label{fig:InterTunneling}  (top panels) ${\bm d}$-dependent tunneling map for an interlayer distance of $3.35$ \AA, focusing here on the real part of the $C_1-B$ interaction. The left panel is obtained from DFT, the middle is obtained by fitting the $p$ and $q$ parameters from the TB model in Eq.~(\ref{STCtunneling}) while the right panel is the difference between the DFT and TB panels. The bottom panel illustrates this fitting for different interactions at different interlayer distances, setting $\exp{\left[{(r_z -v)}/{w}\right]}$ in Eq~(\ref{STCtunneling}).}
\end{figure}

For the inter-layer interactions of our HTC model, we parametrize the TC model introduced in Ref.~[\onlinecite{https://doi.org/10.48550/arxiv.1910.12805}] for interactions between G and h-BN which corrects the conventional Slater-Koster parametrization to accurately capture the tunneling at the K-point at different interlayer distances. This correction can be fitted by an exponential prefactor to the TC expressions as
\begin{equation}
   t^{\rm inter}_{kl} = \exp \left[ \frac{r_z -v}{w} \right]  t^{\rm inter}_{{\rm TC}, kl}.
\label{STCtunneling}
\end{equation}
where
\begin{equation}
    t_{ {\rm TC}, \, kl}^{\rm inter} = V_{pp\pi}(r_{kl}) \left[1 - \left(\frac{r_z}{r_{kl}} \right)^2 \right] + V_{pp\sigma}(r_{kl}) \left(\frac{r_z}{r_{kl}} \right)^2
\label{TCeq}
\end{equation}
with
\begin{equation}
    V_{pp\pi}(r_{kl}) = V_{pp\pi}^0 \exp\left(-\frac{r_{kl}-a_{\rm CC}}{r_0}\right)
    \label{vpppiEq}
\end{equation}
and
\begin{equation}
    V_{pp\sigma}(r_{kl}) = V_{pp\sigma}^0 \exp\left(-\frac{r_{kl}-c_0}{r_0}\right)
\label{vppsigmaEq}
\end{equation}
where $r_{kl}$ is the distance covered by ${\bm r}_{kl}$. 
The tunneling is calculated using the approach outlined in Ref.~[\onlinecite{https://doi.org/10.48550/arxiv.1910.12805}].
The fitted TB tunneling map compared with the original DFT data is shown in Fig.~\ref{fig:InterTunneling} at an interlayer distance of $3.35$ \AA. 
For simplicity, here we have defined a fixed normal vector $\bm r_z$ along the z-axis rather than allowing it to tilt with the local curvature following the surface corrugation. Its associated distance $r_z = 3.35~\AA$ is the interlayer distance, $a_{\rm CC} = 1.42~\AA$ is the rigid graphene's interatomic carbon distance, $V_{pp\pi}^0 = -2.7$~eV the transfer integral between nearest-neighbor atoms, $V_{pp\sigma}^0 = 0.48$~eV the transfer integral between two vertically aligned atoms that were fitted to generalized gradient approximation (GGA) data for fixed interlayer distances~\cite{laissardire2012}.
The decay length of the transfer integral is chosen as $r_0 = 0.184 a_G$ such that the next-nearest intralayer coupling becomes $V_{pp\pi} = 0.1V_{pp\pi0}$.
The fitting parameters entering Eq.~(\ref{STCtunneling}) are given in Table \ref{Table:InterlayerFitting}.
\begin{table}[th]
\begin{center}
\resizebox{0.4\columnwidth}{!}{%
\begin{tabular}{|c|c|c|}
\hline
             &   $v$ (\AA)    &     $w$ (\AA)   \\ \hline
$C_1-B$ & 3.0477  & 1.0280   \\ \hline
$C_1-N$ & 3.4755  & 1.1922   \\ \hline
$C_2-B$ & 3.0646  & 1.0492   \\ \hline
$C_2-N$ & 3.4799  & 1.1622   \\ \hline
\end{tabular}}
\caption{\label{Table:InterlayerFitting} Fitting parameters $v$ and $w$ entering Eq.~(\ref{STCtunneling}) for the  four different possible interactions. The same parameters fit both the real and imaginary part of a specific interaction.}
\end{center}
\end{table}
In Appendix~\ref{Sec:InterTunneling}, we give the DFT to TB tunneling value fitting comparisons for all the other interlayer hopping terms.

\subsection{Truncated atomic plane wave(TAPW) method For average Mass term}
\label{TAPW}
To accurately and efficiently compute the electronic structure of moir\'e systems and extract key quantities such as the average mass term in G/h-BN, we employ the TAPW method~\cite{PhysRevB.107.125112}. This approach provides an optimal balance between the computational efficiency of continuum models and the accuracy of full tight-binding (TB) calculations. The core of the TAPW approach is to project the full TB Hamiltonian onto a reduced subspace spanned by a set of atomic plane waves. These basis functions are constructed as Bloch sums of atomic $p_z$ orbitals, incorporating the real-space positions of all atoms within the moiré unit cell:

\begin{equation}
    |\psi_{\alpha n}(\bar{{\bm k}})\rangle = \frac{1}{\sqrt{N_m N_\alpha}} \sum_{l,i} e^{i(\bar{{\bm k}} + {\bm G}_n) \cdot {\bm R}_{li\alpha}} |\phi({\bm r} - {\bm R}_{li\alpha})\rangle,
    \label{eq:TAPW_basis}
\end{equation}
where $\alpha$ indexes the sublattice and layer, $\bar{{\bm k}}$ is a wavevector in the moiré Brillouin zone (mBZ), ${\bm G}_n$ are the moiré reciprocal lattice vectors, and ${\bm R}_{l i \alpha}$ denotes the position of atom $\alpha$ in the $i$-th atomic cell of the $l$-th moiré cell.

The main approximation, valid for small twist angles, is to retain only those plane waves with wavevectors $\bar{{\bm k}} + {\bm G}_n$ lying within a finite cutoff around the graphene Dirac points (${\bm K}$ and ${\bm K}'$ valleys). This truncation respects the approximate valley $U(1)$ symmetry and allows to reduce the Hamiltonian dimension while preserving the low-energy band structure with high fidelity.

The projected Hamiltonian is expressed as
\begin{equation}
    H^{TAPW} = {\bm X}^\dagger {\bm T} {\bm X}
\label{TAPW_X}
\end{equation}
where ${\bm T}$ is the full TB matrix based on our hybrid TC (HTC) model described in Sec.~\ref{TBSect} containing the hopping integrals $t({\bm r}_{i\alpha, j\beta})$, and ${\bm X}$ is the plane wave projection matrix with elements $({\bm X}_\alpha)_{n, i} = e^{i{\bm G}_n \cdot \boldsymbol{\tau}_{i\alpha}}/\sqrt{N_\alpha}$. This formulation provides a mapping between the atomistic TB description and a generalized continuum model.

In our study of G/h-BN heterostructures, the sublattice-asymmetric potential induced by the h-BN substrate is encoded in the on-site energy term $\Delta_\alpha$ within the TB Hamiltonian ${\bm T}$. This term differs for the two graphene sublattices (A and B) due to their registry with the underlying boron and nitrogen atoms. The average mass term $\bar{m}$ that contributes to the primary band gap at the Dirac point is then directly computed from the projected low-energy Hamiltonian as
\begin{equation}
    \bar{m} = \frac{\Delta_A - \Delta_B}{2}.
    \label{eq:avg_mass}
\end{equation}
By employing the TAPW method, we efficiently and reliably compute the substrate-induced average mass term across a wide range of twist angles and under different structural relaxation schemes. This allows us to systematically establish its dominant role in the formation and modulation of the primary band gap in G/h-BN moiré heterostructures.

\subsection{Lattice reconstruction after energy minimization calculations in G/h-BN}
\label{MDSection}
To study the relaxation effects in our different configurations, we perform energy minimizations using the \texttt{LAMMPS} package~\cite{PLIMPTON19951}.
For the interlayer interaction force-fields, we use the DRIP potential~\cite{PhysRevB.98.235404} that has been reparametrized~\cite{https://doi.org/10.48550/arxiv.1910.12805} using EXX-RPA DFT calculations~\cite{PhysRevB.96.195431}
which includes dihedral angle corrections to improve on the well-known registry-dependent Kolmogorov-Crespi (KC) potential~\cite{Kolmogorov_2005}. 
For the intralayer interactions, we use the extended Tersoff potential (ExTeP)~\cite{PhysRevB.96.184108} for the B-N interactions and the REBO2 potential~\cite{Brenner_2002} for the C-C interactions.

\section{Results}
\label{sec:Result}
\begin{figure}[tbhp]
\centering
\includegraphics[width=1.0\columnwidth]{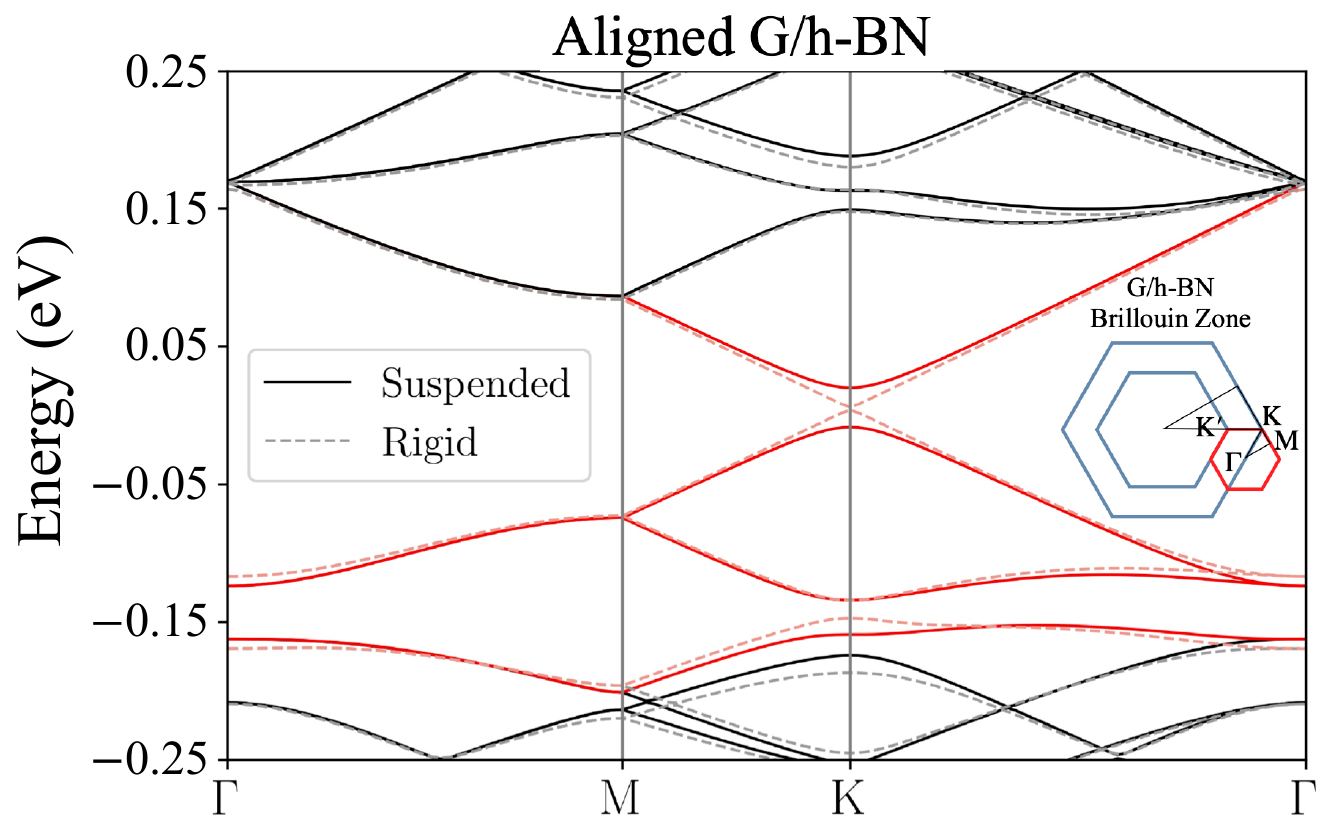}
 \caption{\label{fig:GBNBand} Electronic band structure for the rigid configuration (dashed lines) and the suspended configuration (solid lines). The increase in primary gap and decrease in secondary gap due to relaxation effects is clearly visible, where the primary gap goes from $3.51$ meV to $20.3$ meV and the secondary gap from $16.41$ meV to $14.86$ meV. The inset shows the Brillouin Zone for G/h-BN with a red hexagon indicating the $\Gamma-M-K-\Gamma$ along which the band structures are calculated. Here we set to zero the Fermi level inside the gap by shifting the bands by $-0.73$ eV.}
\end{figure}
We first remind in Fig. \ref{fig:GBNBand} the well-known qualitative effect that relaxation effects have on the electronic band structure of the \textit{suspended G/h-BN} geometry (solid line) when compared to the rigid band structure (dashed line). The moir\'e BZ created by the G/h-BN superlattice with the relevant high-symmetry k-path is given in the inset, where the larger hexagon corresponds to the BZ
of the graphene layer,
the smaller hexagon to the BZ of the h-BN layer,
We observe the expected gap opening for the primary and secondary hole-side Dirac point, matching the behavior observed in existing literature~\cite{10.1021/acs.nanolett.8b03423, jung2013_PRB}. The main effect of the relaxation is to increase the amplitude of the primary gap and to decrease the size of the secondary gap.
We note here that the primary and secondary band gaps in graphene due to aligned h-BN substrates can be quite sensitive to the choice of the atomic force fields and electronic structure models due to varying inter-layer distances and changing coupling strengths. 
In Appendix~\ref{AppendixD}, we present additional band structures for different twist angles of G/h-BN.
\begin{figure*}[tbhp]
\centering
\includegraphics[width=0.9\textwidth]{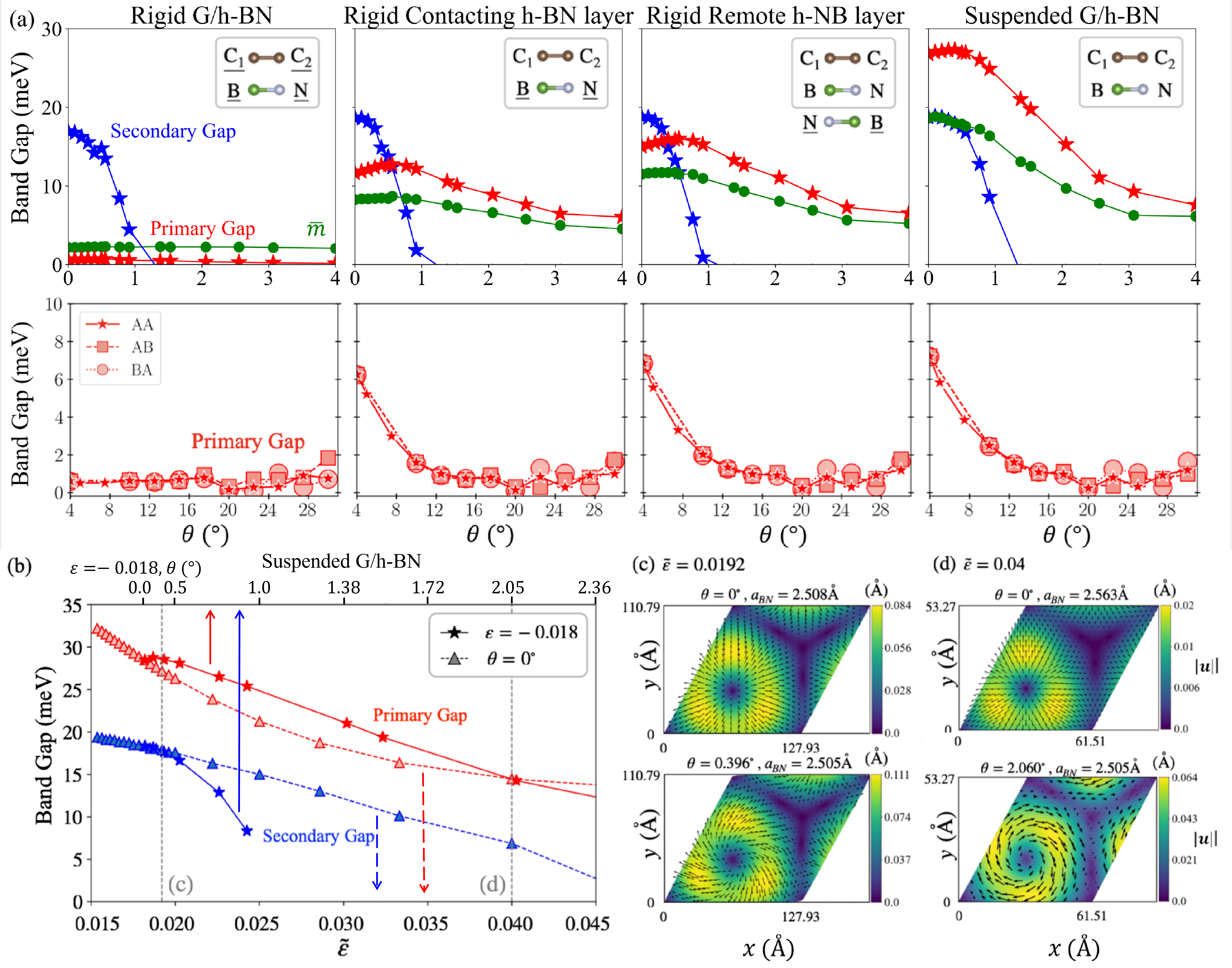} 
\caption{\label{fig:Angle} (a) Primary (red) and secondary (blue) gap estimates and average mass term $\overline{m}$ (green) for G/h-BN with respect to the twist angle $\theta$.  In the upper panels, we show the small angle behavior which shows a strong reduction by a factor 2 to 4 from the {suspended G/h-BN} configuration when a substrate is included. A small inflection point near $0.5^\circ$ is also seen for these substrate configurations. The sketches indicate which atoms are kept rigid during the relaxation by labeling them with underline and the grey box shows which atoms are included in the band structure calculation. In the lower panels, we show the large angle behavior illustrating the robust primary gap up to $30^\circ$ which becomes dependent on the rotation center being at the $AA$, $AB$, or $BA$-stacking (star, square, or circle, respectively).
(b) shows the angle-dependence of the bandgaps similar to (a) for the {suspended G/h-BN} configuration but in terms of an effective lattice constant mismatch $\tilde{\epsilon}$ that gives the same moire period, see Eq.(~\ref{effectiveEps}), and illustrates the role of including rotations in the Hamiltonian. 
The quiver plots in (c) and (d) for suspended G/h-BN geometries considered show that the local displacement amplitudes $|{\bm u}|$ are larger in rotated geometries up to three times for systems with same moire periods constructed by modifying the lattice mismatch $\varepsilon$ but with $\theta = 0^{\circ}$.}
\end{figure*}

Our main results are summarized in Fig.~\ref{fig:Angle}
where we consider the angle-dependence of the primary and secondary gap estimates. For this, we consider 4 different configurations, namely \textit{suspended G/h-BN} mentioned above where the two layers are fully relaxed during the classical force field calculations, \textit{rigid contacting h-BN layer} where the interface h-BN layer is kept rigid during relaxation to mimic the effect of a substrate at a first level of approximation, \textit{rigid remote h-NB layer} where we add an additional remote h-BN layer at a $180^\circ$ rotation (h-NB) and fix its atomic positions, to allow the relaxation of the interface h-BN layer, and the \textit{rigid G/h-BN} where none of the layers are allowed to relax. The layers that are kept fixed during the energy minimization are indicated by an underline in the insets of Fig.~\ref{fig:Angle}(a). For a more accurate study of the substrate effect, one could introduce several tens of layers~\cite{10.1021/acsnano.9b00645} but this approach is beyond the scope of this study.

We observe that for all configurations, the secondary gap closes at about $1.0^{\circ}$ twist angle as shown in Fig.~\ref{fig:Angle} with small differences in their respective amplitudes. The amplitude of the primary gap however is quite sensitive to the specific relaxation conditions where its angle-dependent value remains almost flat in rigid G/h-BN 
while it decreases gradually for the other three configurations. The largest gap predictions happen when all layers are allowed to relax while they get damped when the effect of a substrate is considered. 
In the presence of a substrate, we notice the presence of a non-monotonic behavior showing a weak increase of the primary gap around $~0.5^\circ$, while this effect is nearly absent when fully relaxing the system. 

The average effective inter-sublattice potential difference within graphene can be estimated from the average mass term $\overline{m}$ in Eq.~(\ref{eq:avg_mass}) through the TAPW method.
The evolution of this average mass-term contribution follows closely the band gap and accounts for more than 70~\% of its magnitude.
\begin{figure*}[tbhp]
\centering
\includegraphics[width=0.9\textwidth]{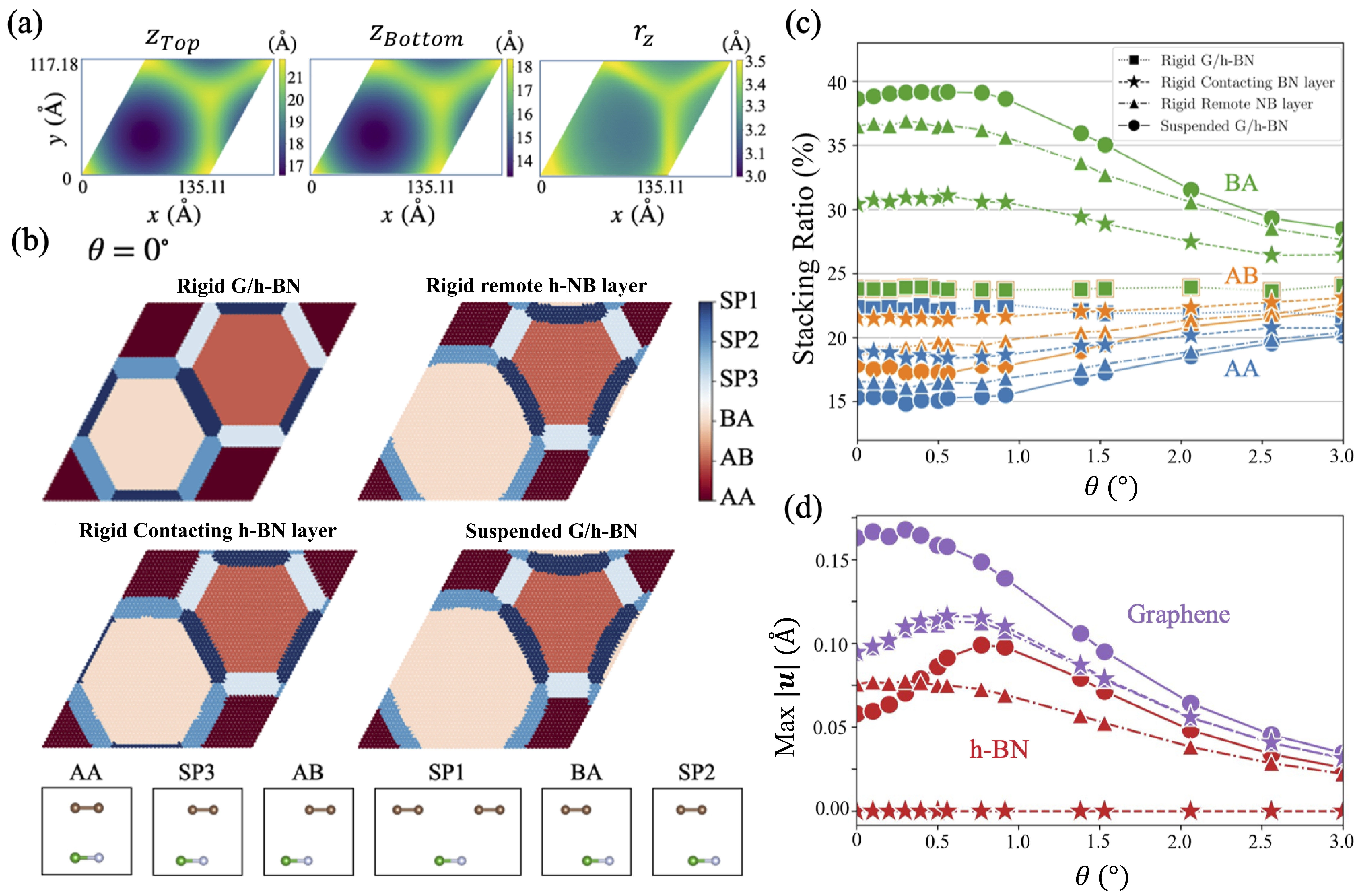}
\caption{\label{fig:RelaxedStructure}
(a) z-coordinate of the atoms in the top and bottom layers as well as their subtraction illustrating the interlayer distance for the suspended geometry indicating that the most stable BA-stacking configuration gives the smallest interlayer distance. 
(b) Assignment of each of the atoms in the top layer to one of the seven stacking configurations from the bottom sketches based on 
the value of ${\bm d}_{ij}$ using the approach outlined in
Ref.~[\onlinecite{https://doi.org/10.48550/arxiv.1910.12805}]. (c) Angle-dependence of the relative stacking ratios for the AA, AB and BA stacking regions from (b). (d) Maximum value among the all the displacements of the atoms after relaxation with respect to their position before relaxation $\max(|{\bm u}|)$ using the matching symbols from panel (d) where the purple curves correspond are extracted from the graphene layer atoms and the red curves come from the h-BN layer atoms. The maxima in (c) and (d) match the larger primary gap reported in Fig.~\ref{fig:Angle}(a) suggesting that the larger lattice reconstruction away from zero alignment when a substrate is present causes the increases the primary gap size.  
}
\end{figure*}
To explain the observations, including the small-angle inflection point that has been reported but left unexplained in Ref.~[\onlinecite{PhysRevB.100.195413}],
we quantify in Fig.~\ref{fig:RelaxedStructure} the strength of the relaxation effects for the aligned G/h-BN geometry using the approach outlined in Ref.~[\onlinecite{https://doi.org/10.48550/arxiv.1910.12805}] to assign each atom to one of the main highly symmetric stacking configurations (AA, AB, BA, SP$_{1}$, SP$_{2}$ and SP$_{3}$) based on the components of its sliding vector ${\bm d}$. A more robust distance-dependent-only recipe was introduced in Ref.~\cite{PhysRevB.110.155419}. This approach allows us to transition from panel (a) where the interlayer distance map after relaxation shows a continuous distribution of possible stacking configurations in the moir\'e pattern to panel (b) where we observe a discrete distribution of stacking configurations. In the top left figure of (b), we show the stacking distribution for the rigid structure and in the other panels of (b) we illustrate this distribution for the different relaxed geometries considered here. We note that after relaxation the AA and AB stacking regions decrease in size and the BA stacking region increases in size as expected from energetic considerations where the BA stacking is most stable. In panel (c) we then show the ratios that each of these configurations occupy in the moir\'e system for varying twist angles. In panel (d) we finally illustrate the absolute value of $|{\bm u}|$ which captures by how much an atom moves from its rigid position due to lattice reconstruction.
The trends observed in (c) and (d) match the observations made in Fig.~\ref{fig:Angle}, \textit{i.e.} the \textit{suspended G/h-BN} geometry with the largest lattice reconstruction presents the largest primary gap estimates. Furthermore, for all configurations other than the rigid, we observe that the inflection point at small twist angle around~$0.6^\circ$ is in keeping with the relatively larger BA stacking area. The rigid configuration shows nearly the same primary gap estimates for all angles since lattice reconstruction is absent by definition.

\begin{figure}[tbhp]
\centering
\includegraphics[width=1.0\columnwidth]{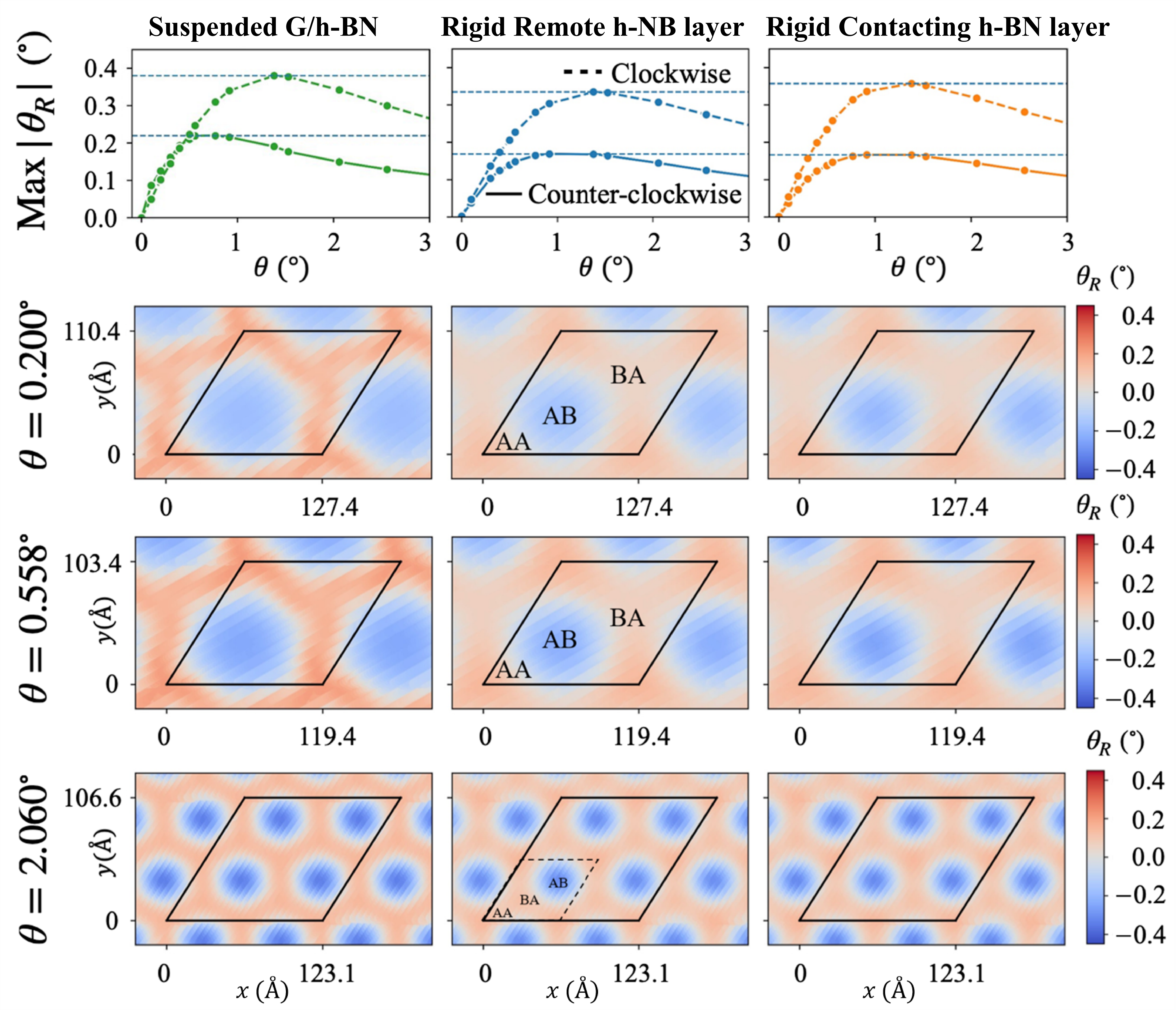} 
\caption{\label{ThetaR} (top row) Maximum value from the twist angle $\theta$-dependent local rotation angle $\theta_R$ at the center of the AA (counter-clockwise) and AB (clockwise) stacking regions for three different relaxation schemes. (other rows) $\theta_R$ for three different values of $\theta$, namely $0.2^\circ$, $0.558^\circ$ and $2.06^\circ$ for the same relaxation schemes. The substrate relaxation schemes dampen the local rotation angles. These plots indicate that the larger displacements in Fig.~\ref{fig:RelaxedStructure}(c) and (d) root in stronger rotation effects that stabilize the system and increase the primary gap size.}
\end{figure}

To investigate in greater detail the enhanced lattice reconstruction observed at small twist angles, we examine in Fig.~\ref{ThetaR} the evolution of the local rotation angle $\theta_R$ for different twist angles at selected high-symmetry stackings, together with representative maps showing its spatial distribution across the moiré unit cell. Rather than adopting the continuum definition of $\theta_R$ from Ref.~[\onlinecite{Kazmierczak2021}], we define it directly in real space as the rotation angle between a given dimer before and after relaxation (see Appendix for illustration).
Interestingly, as shown in Fig.~\ref{thetaRFig} of Appendix~\ref{thetaRSec}, the resulting $\theta_R$ values depend on which of the three inequivalent A-sublattice–centered dimers is used to construct the map. This choice breaks the nominal triangular symmetry of the moiré lattice. By averaging over the three dimer orientations, the expected symmetry is restored, yielding the maps shown in panel~(b) of the current figure.
Panel~(a) reveals that the local rotation angle at the AA and AB stackings evolves in the opposite sense to the global twist, whereas at the BA stacking it reinforces it. This contrast becomes more pronounced at small but finite twist angles, consistent with the stronger lattice reconstruction seen in Fig.~\ref{fig:RelaxedStructure}.

To further emphasize the significance of these local rotation effects, panel~(b) of Fig.~\ref{fig:Angle} compares the twist-angle dependence of the band gaps against a model with zero twist angle but equivalent lattice mismatch to give the same moire period valid for small $\theta$,~\cite{Jung2015} expressed as
\begin{equation}
\tilde{\epsilon} = \sqrt{\epsilon^2 + \theta^2},
\label{effectiveEps}
\end{equation}
where $\epsilon = \alpha - 1$ is the lattice mismatch between $a_G$ and $a_{BN}$.
We find that the gap magnitudes are comparable for $\tilde{\epsilon} = 0.0192$ corresponding to $\theta = 0^\circ$, they differ markedly at $\tilde{\epsilon} = 0.040$ corresponding to $\theta = 2.06^\circ$. The latter case exhibits strong local rotation displacements, as visualized in the quiver plots of Figs.~\ref{fig:Angle}(c) and (d), where the arrow directions indicate the displacement vector ${\bm u}$ and the colormap encodes its amplitude. 

Finally, as shown in the lower panels of Fig.~\ref{fig:Angle}(a), our simulations predict that the primary gap remains finite at large twist angles, saturating around $\sim 1$~meV up to $30^\circ$. This behavior contrasts with continuum-model predictions of a complete gap closure at large angle.~\cite{10.1103/physrevb.96.085442} At these higher angles, the calculated gap becomes sensitive to the specific rotation center used to construct the commensurate cell. While this choice is immaterial at small angles, where the large moiré period averages over all local stacking configurations, the transition to a discrete set of local stackings near $\sim 15^\circ$ renders the gap values rotation-center dependent. This discretization effect is reminiscent of the even–odd distinction known to yield distinct band structures near $30^\circ$ rotation.~\cite{PhysRevB.81.161405}

We further calculate the sign associated with the band-gap the twist angle and observe that no sign change happens between different twist angles up to $30^\circ$, see Appendix~\ref{AppendixB}. This effectively means that for all angles up to $30^\circ$ the wavefunction will preferably be located on the same carbon sublattice without changing sublattice when twisting the graphene layer. Beyond this value, due to the inequivalent B and N atoms and the trigonal symmetry of the lattice, the wavefunction distribution flips from one carbon sublattice to the other carbon sublattice.
We finally examine the total energy as a function of twist angle. For the rigid configuration, the energy remains essentially independent of angle, indicating the absence of an intrinsic rotational preference in the absence of relaxation. In contrast, for the suspended configuration, the energy decreases as the system approaches alignment, showing a clear tendency to rotate back toward the aligned state. When a substrate is present, and consistent with the enhanced lattice reconstruction observed at small but finite twist angles, the system instead exhibits a tendency to lock near $0.6^\circ$.
In finite flakes, such locking has been attributed to the alignment of the moir\'e lattice vectors with the flake edges happening at an analytically predicted angle of $0.61^\circ$ for armchair-terminated flakes~\cite{2510.18694}. In the bulk limit considered here, however, the relevant alignment occurs between the moir\'e lattice vectors and the graphene lattice vectors, happening at the same twist angle. This bulk alignment mechanism is likely the dominant factor for microscopic flakes~\cite{doi:10.1126/science.aad2102, Woods2016,Ribeiro_Palau_2018,doi:10.1021/acsami.3c00558,PhysRevB.94.045401,Silva2020-ex}.

\begin{figure}[tbhp]
\centering
\includegraphics[width=0.9\columnwidth]{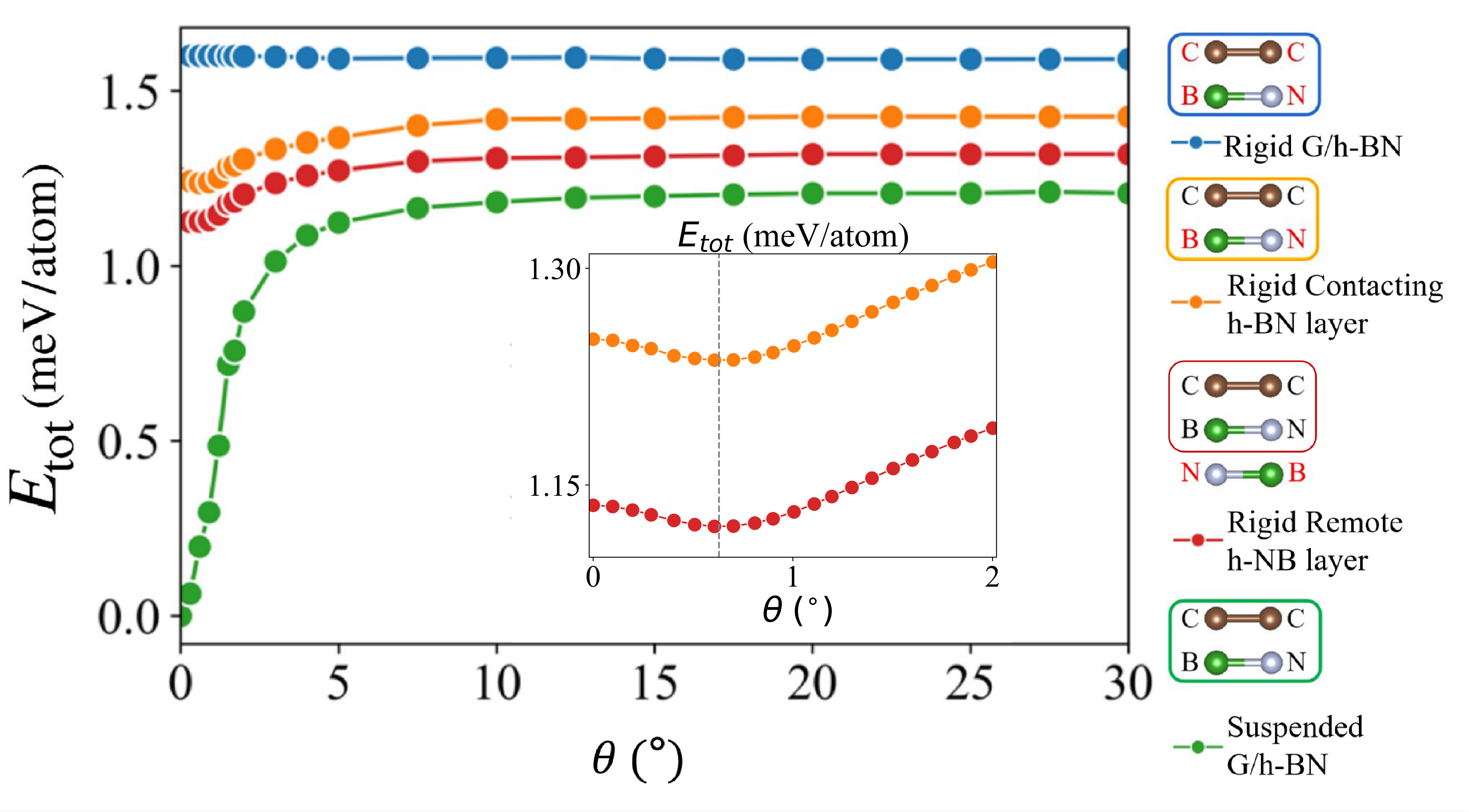}
\caption{\label{TotalEnergy} Energies renormalized per atom for for three different relaxation types. The values are globally shifted by 
$-7.07020$~eV/atom to bring the \textit{suspended G/h-BN} zero aligned configuration to zero. For rigid remote NB layer case, we shifted the energy by an extra 111.68 meV/atom to get rid of the substrate h-NB layer energy. We zoom in on angles near $0.6^\circ$ in the inset. These curves illustrate that the stronger lattice reconstructions near this angle from Figs.~\ref{fig:RelaxedStructure}~and~\ref{ThetaR} lead to weak energetic stabilization of the systems around this same angle. Beyond the range illustrated in the inset, no preferential angle values are observed. The simulation cells are based on the commensurate cells from Table~\ref{Commensurate_2} in Appendix~\ref{energyAppendixSect}.
}
\end{figure}

\section{Conclusion}

Using atomistic energy minimization simulations we show that the presence of a substrate affects quantitatively the expected primary and secondary gap reduction when twisting the graphene layer away from alignment with respect to the h-BN layer. With the force-fields used in this paper, the primary gap is reduced by a factor two or four depending on which substrate approximation is used, illustrating their importance in how graphene-based moir\'e superlattices are modeled. We speculate that using a computationally more expensive but more accurate GAP$_{20}$~\cite{Rowe2020} force field in terms of the elastic constants of graphene, leading to increased stiffness of the system, one can expect a further reduction of the gap estimates. The secondary gap is less sensitive to the choice in relaxation scheme and drops from a value close to $12$ meV down to zero for twist angles beyond $1^\circ$.  The primary gap is robust up for all angles up to $30^\circ$, and the discreteness of local stacking contributions activates the rotation center degree of freedom for systems with twist angles larger than $20^\circ$, affecting the numerical values of the observed gaps. The small increase of the order of a few meV in the primary gap at $0.6^\circ$ roots in a larger lattice reconstruction as compared to the aligned geometry, thus suggesting that it is possible to stabilize a non-zero twist angle provided that the corrugation of the interface h-BN can be suppressed.

\acknowledgements
This work was supported by the Korean NRF through Grant NRF-2020R1A5A1016518 (J.J.) and Grant RS-2023-00249414 (N.L.).
We acknowledge computational support from KISTI Grant No. KSC-2022-CRE-0514 and by the resources of Urban Big data and AI Institute (UBAI) at UOS.

\newpage
\bibliography{all}

\clearpage
\appendix

\section{TB model and electronic structure considerations}

\subsection{Bilayer vs single layer F2G2}
\label{AppendixF2G2}

Here we illustrate the agreement between the DFT and the TB models for the high symmetry stacking configurations of 4-atom commensurate G/h-BN unit cells, focusing solely on the intralayer term effects. We use either the F2G2 models for Graphene and h-BN extracted from single layer calculations or we use the F2G2 model extracted from a bilayer G/h-BN calculation as presented in the main text. The interlayer terms are kept unchanged from the DFT calculation. We see that the bilayer F2G2 model gives the best agreement with DFT while the single layer F2G2 gives reasonable agreement at low energy and this would probably constitute a reasonable first approximation for more complex layered Graphene and h-BN combinations.

\begin{figure}[tbhp]
\centering
\includegraphics[width=1.0\columnwidth]{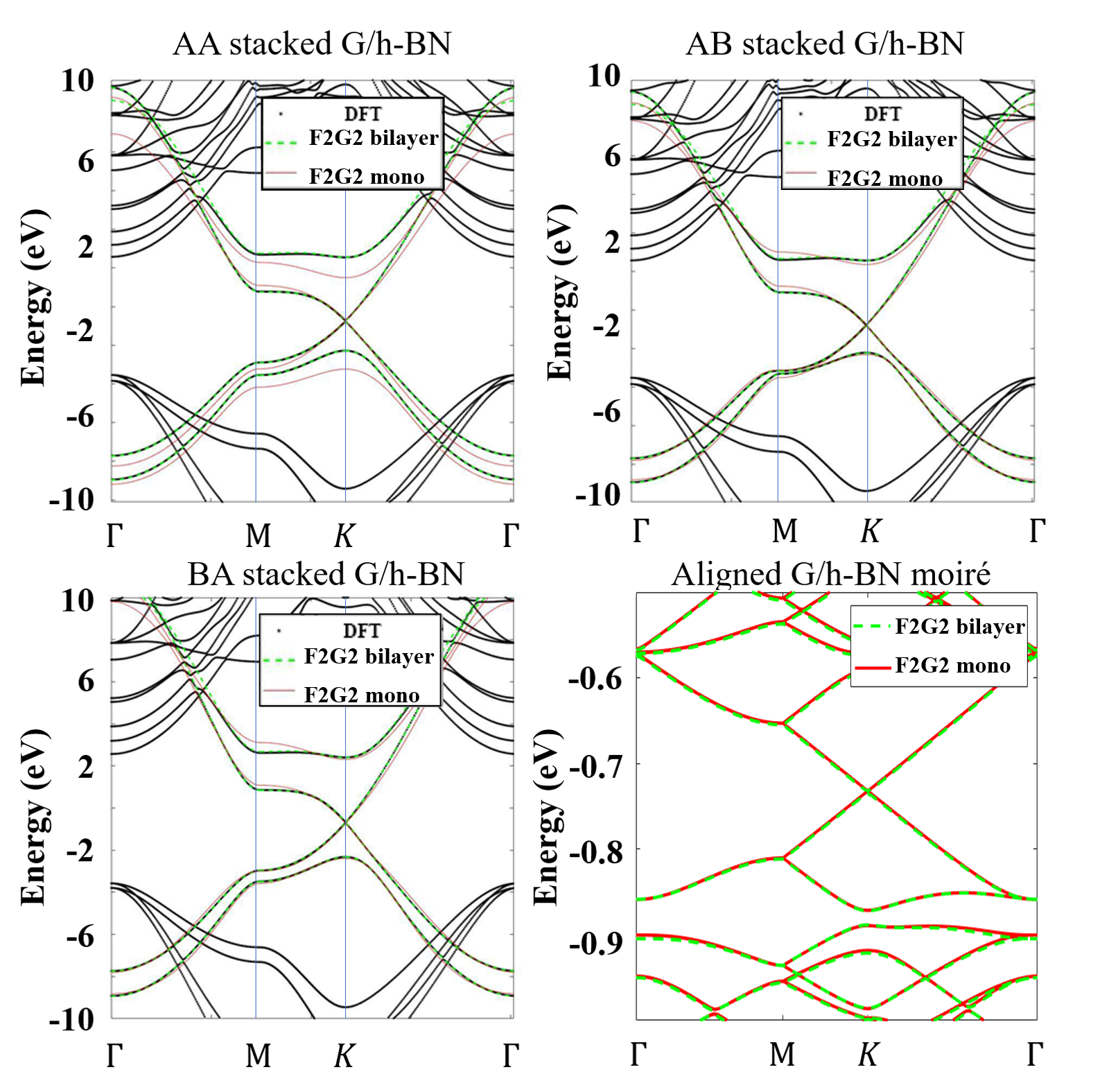}
\caption{\label{F2G2Fig} Electronic bandstructure comparisons between full DFT calculations and TB model where the intralayer terms use either monolayer Graphene and h-BN parametrizations or a bilayer G/h-BN parametrization while the interlayer terms are given by the DFT panels for the first 3 panels for the 4-atom commensurate cell curves and by our reparametrized TC model in the last panel on aligned G/h-BN. The HTC model used in this paper would then correspond to the dashed green curve in this panel. We note that the bilayer F2G2 reproduces the 4-atom cells best while the single F2G2 predicts for the aligned moire system a smaller primary gap value, from $3.5$~meV to $0.3$ meV, while the secondary gap remains nearly unchanged, from $16.4$ meV to $17.8$ meV, for bilayer and single intralayer terms, respectively. 
}
\end{figure}

\subsection{Interlayer-distance dependent intralayer moir\'e terms}
\label{AppendixA}
\begin{figure}[tbhp]
\centering
\includegraphics[width=1.0\columnwidth]{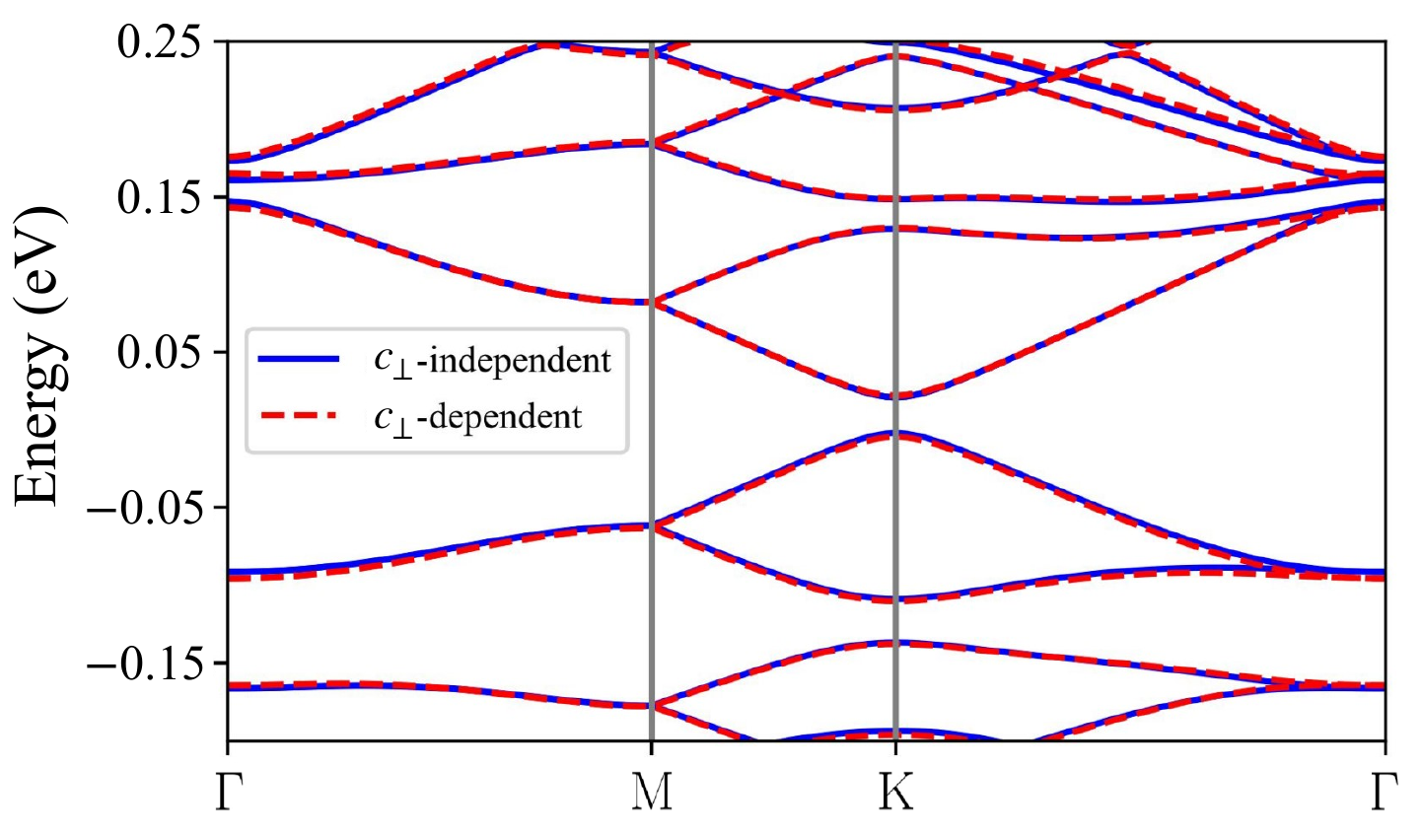}
\caption{\label{cperp} Comparison between the band structures of aligned h-BN when the interlayer-distance dependent parametrization of the DFT calculation is considered in our intralayer Hamiltonaion terms and when a fixed interlayer distance of $3.35$ \AA\ is assumed. This dependence is ignored in the calculations from the main text.}
\end{figure}

In the figures presented in the manuscript, we used parametrizations for the intra-layer terms that assume a fixed interlayer distance of $3.35$~\AA\ when performing the DFT calculations. While we included the changes in the inter-layer hopping terms in the presence of corrugations, we also provide a parametrization of the intra-layer moire potentials in Eq.~(\ref{quadEq}) with interlayer distance. Here, in Fig.~\ref{cperp} we illustrate the negligible corrections to the bands that are introduced by the intra-layer terms due to varying inter-layer distance.

\subsection{Inter-layer Tunneling Maps}
\label{Sec:InterTunneling}
\begin{figure}[tbhp]
\centering
\includegraphics[width=1.0\linewidth]{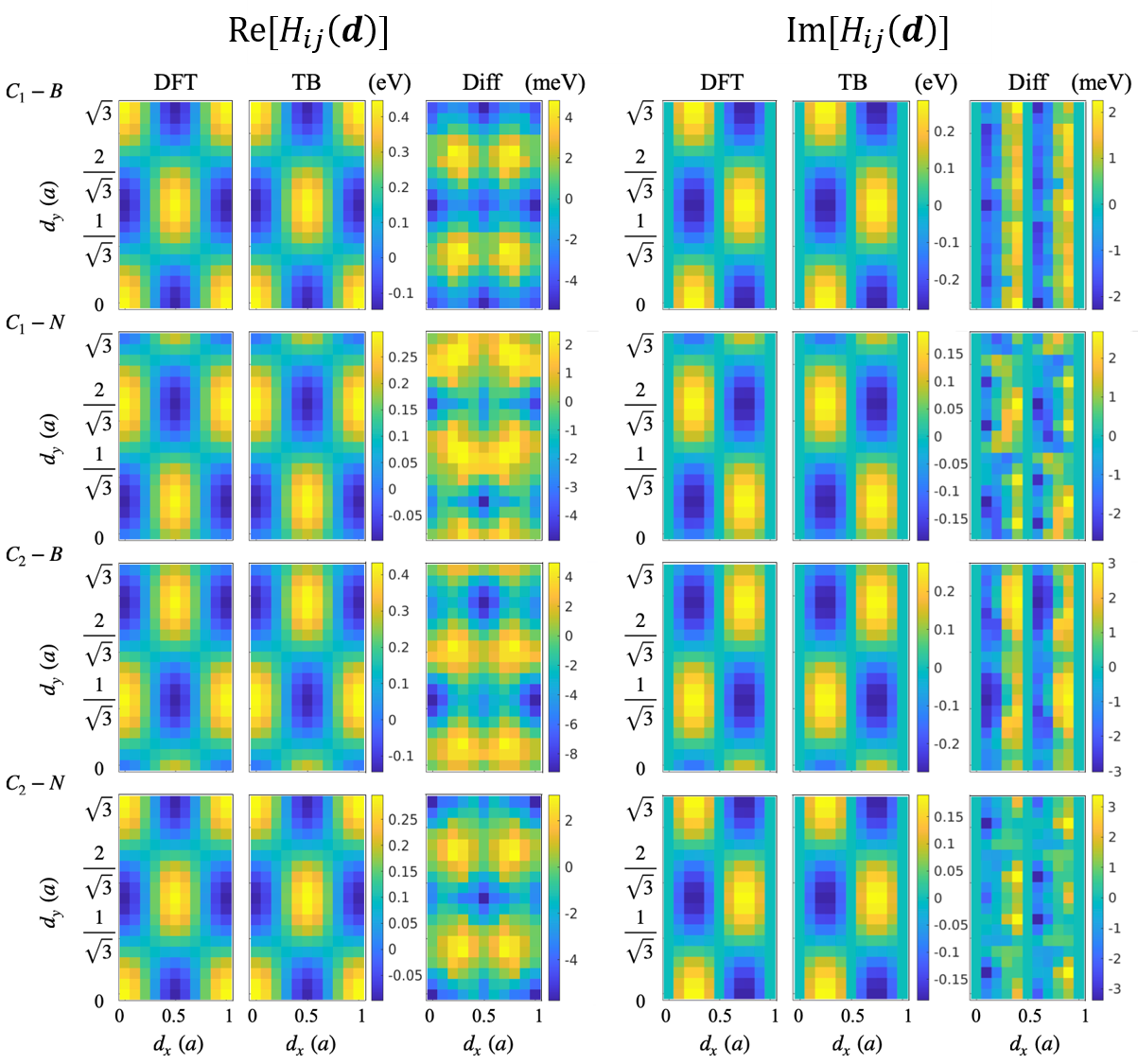}
\caption{\label{fig:MapInterlayerAll} Additional comparisons between DFT and TB tunneling maps as first shown in Fig.~\ref{STCtunneling} for $C_1-B$ repeated in the first row here. The other rows show the same order of magnitude in the difference between DFT and TB for $C_1-N$, $C_2-B$ and $C_2-N$ interactions.}
\end{figure}

We illustrate in Fig.~\ref{fig:MapInterlayerAll} how the DFT and the fitted TB model tunneling displacement maps for the other matrix elements give equally good agreement as was shown in the main text in Fig.\ref{fig:InterTunneling}. We further illustrate that the fitting parameter works equally well for the real and imaginary parts.

\begin{figure}[tbhp]
\centering
\includegraphics[width=1.0\linewidth]{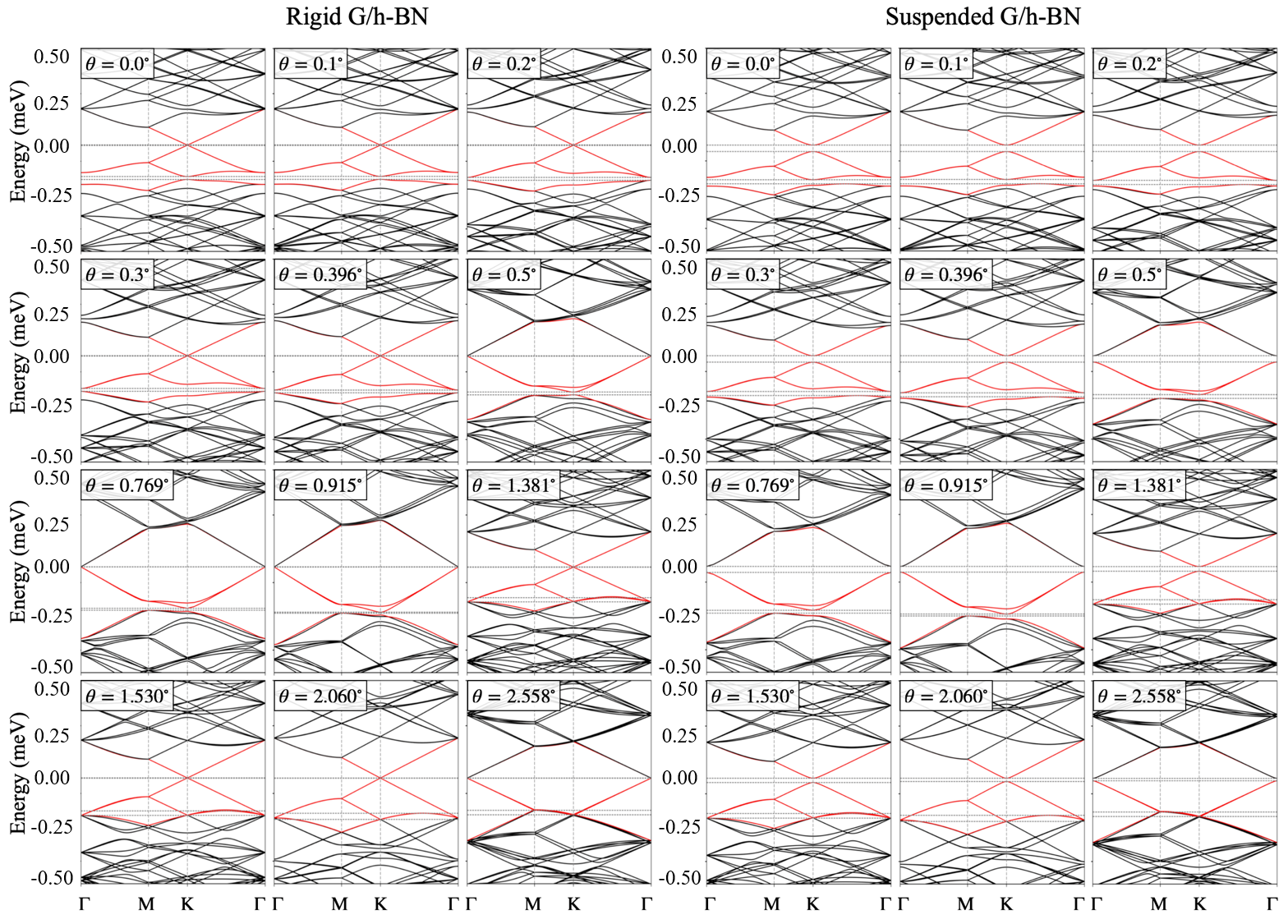}
\caption{\label{fig:BandAll} Band structure of rigid G/h-BN and \textit{suspended G/h-BN} for different twisted angles. The horizontal lines define the edges of the bands that define the primary and secondary gap on the hole side. Those bands are highlighted in red.}
\label{appendixBands}
\end{figure}
\subsection{Electronic band structures}
\label{AppendixD}
For reference purposes, we provide in Fig.~\ref{appendixBands} the band structures for the rigid and the \textit{suspended G/h-BN} configurations that provide us with the gap estimates reported in the paper. Depending on the size of the super-cell, the Dirac point is folded back onto the $\Gamma$-point, e.g. for $0.558^\circ$.

\subsection{Probability distribution of band gap edge states}
\label{AppendixB}

\begin{figure}[tbhp]
\centering
\includegraphics[width=1.0\linewidth]{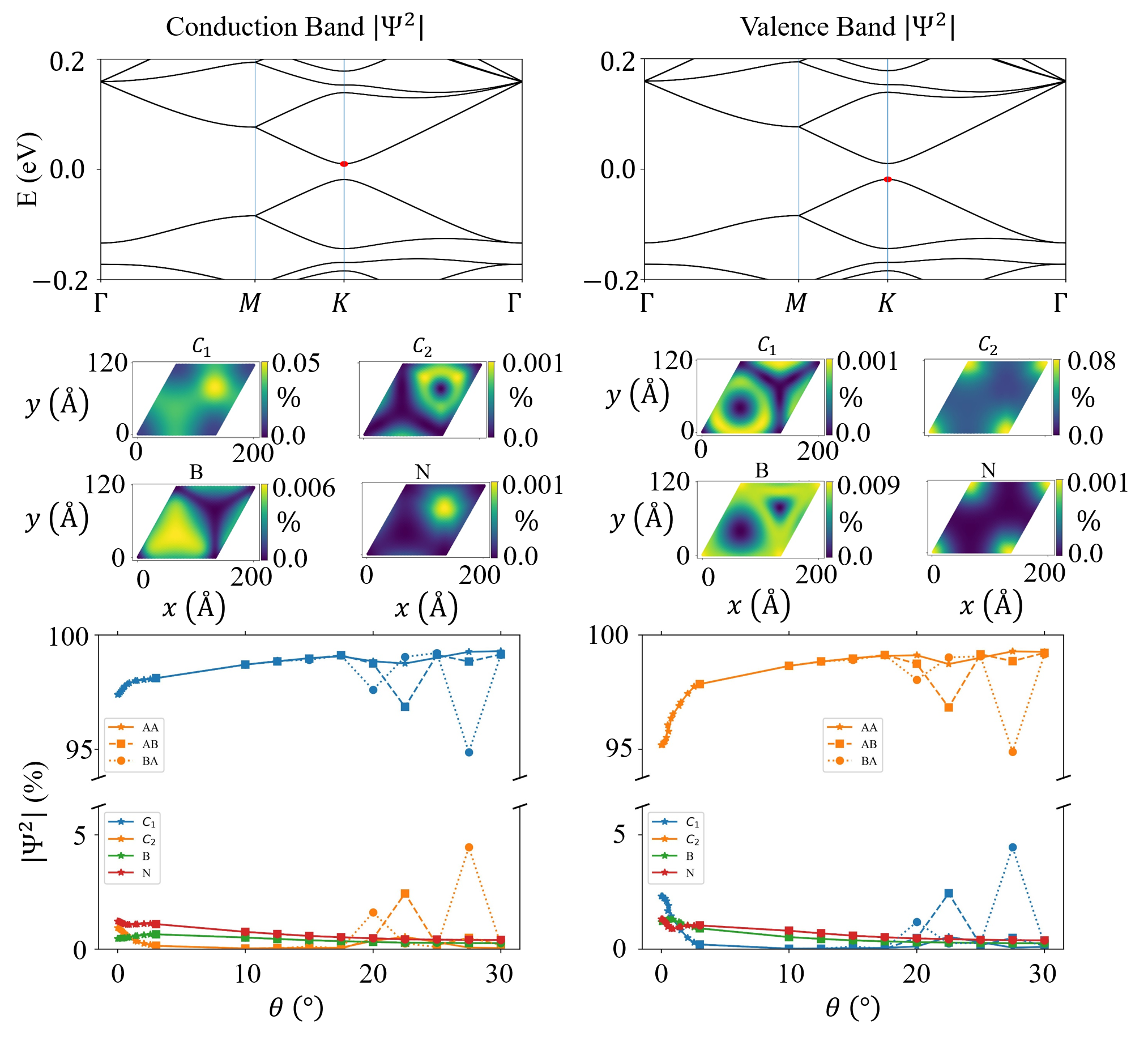}
\caption{\label{fig:Wavefunction} Probability distribution maps for aligned \textit{suspended G/h-BN} for states at both the valence and conduction band edges of the bandgap as well as their angle dependence where the states remain located on the same sublattices regardless of the specific angle value. The bandgap states from the valence band are preferentially located on the $C_1$ sublattice while the valence band states prefer the $C_2$ sublattice. For the mirror angles $60^\circ-\theta$ (not shown here), the role of $C_1$ and $C_2$ appears flipped with respect to $\theta$ angles.}
\end{figure}

In the main text, we noted that the band gap varies with twist angle for the G/h-BN configurations considered here. A promising avenue when such gaps occur is to check if the sign associated with each of these angle-dependent gaps flips between angles. However, this does not seem to be the case. We studied this behavior by looking at the distribution of the wave functions and associated electronic states corresponding to the eigenenergies forming the band edges of the band gap as illustrated here in Fig.~\ref{fig:Wavefunction}. 
By focusing on \textit{suspended G/h-BN}, we illustrate the probabality distribution of each type of atom forming the layered system, differentiating between both C-sublattices (referred to as $C_1$ and $C_2$) as well as the $B$ and $N$ atoms of the h-BN layer. From the colormaps, we observe that the probability distribution is always larger for the graphene layer in both band edges occupying the $C_2$-sublattice for the conduction band the $C_1$-sublattice for the valence band with nearly zero contributions on either the $B$ and $N$ atoms for both the conduction and valence band edges.
These observations remain valid when changing the twist angles in the lower panels thus showing that no sign change in the band gap occurs between angles. The choice in rotation center also does not affect this conclusion. We note that due to the symmetry breaking caused by the inequivalent B and N atoms, the sign flips for all complementary $60-\theta$ angles, thus flipping the preferential sublattice on which the valence band or sublattice bands will be located when compared to their $\theta$ counterparts.

\section{Local rotation angle $\theta_R$}
\label{thetaRSec}

\begin{figure}[tbhp]
\centering
\includegraphics[width=1.0\columnwidth]{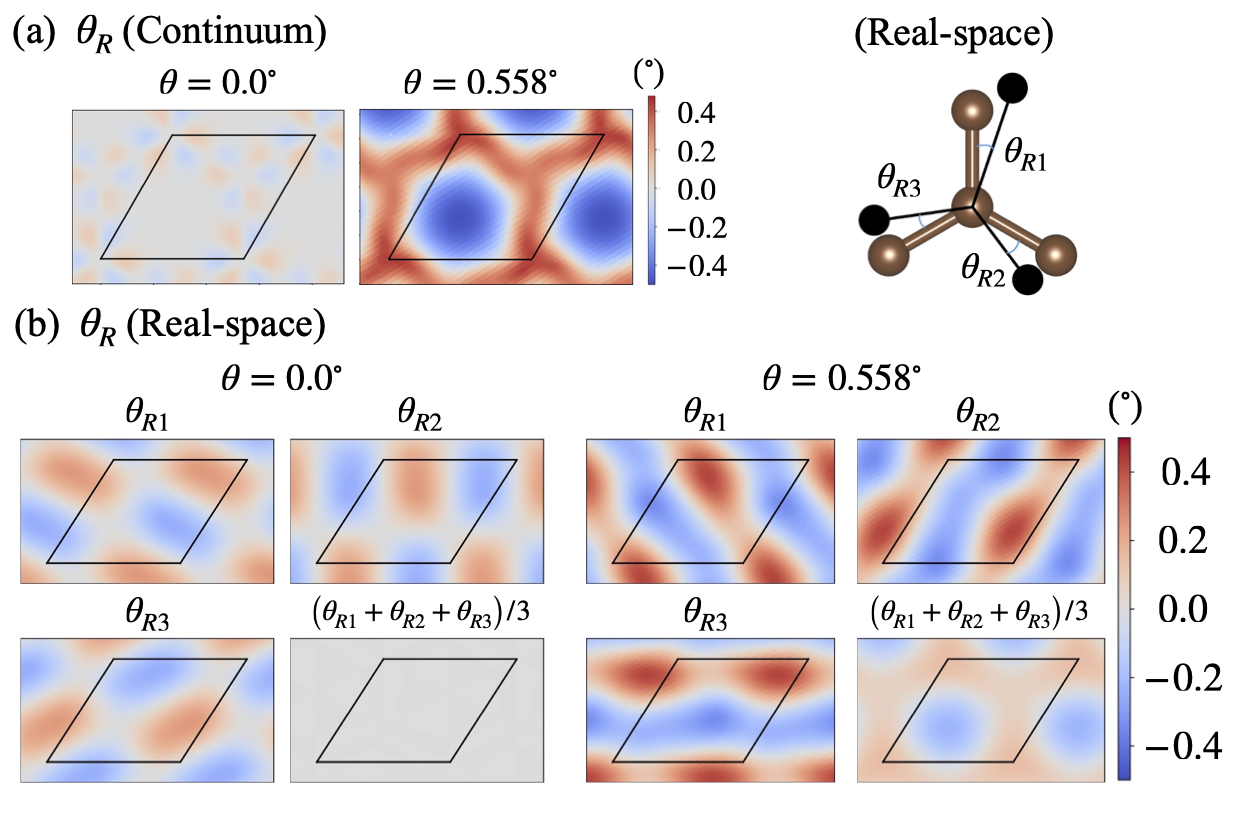}
\caption{\label{thetaRFig} (color online) (a) Continuum and (b) real space calculation of the local rotation angles for $\theta=0^\circ$ and $\theta=0.558^\circ$ where for the real-space we use each of 3 possible choices to define the dimer around a reference atom (see sketch). The last panel of each real-space calculation shows the average of the 3 previous panels where near-perfect cancellation is observed at zero twist. We note overall qualitative agreement of the average real-space approach with the corresponding panels in (a) extracted using Eq.~\ref{contThetaR} based on Ref.~[\onlinecite{Kazmierczak2021}] albeit with differences in the strengths.} 
\end{figure}

In Fig.~\ref{thetaRFig}, we illustrate the qualitative and quantitative differences and similarities between the local rotation angle calculated using the continuum model expression~\cite{Bediako2020} 
\begin{equation}
 \theta_R = s_{yx} - s_{xy} 
 \label{contThetaR}
\end{equation}
where
\begin{equation}
s_{xy} = \frac{\partial u_x}{\partial y}
\end{equation}
and
\begin{equation}
s_{yx} = \frac{\partial u_y}{\partial x}
\end{equation}
where $s_{xy}$ and $s_{yx}$ are simple shear strains calculated from the in-plane displacement vector components ${\bm u} = (u_x, u_y)$
and the real-space approach where we calculate the angle formed by the dimer before and after relaxation (see sketch). We note that the real-space approach gives different results depending on which of the 3 dimers connected to one reference atom is chosen. For the aligned case, we see that when taking the average of $\theta_R$ obtained for the 3 dimer choices, the local rotation for each atom nearly equals zero, as expected from the continuum model expressions. However, the individual dimer contributions show that individual dimers rotate sizeably. A similar dimer-resolved behavior is observed for the rotated case, as illustrated for $\theta=0.558^\circ$. The average value gives the expected counter-clockwise rotation at the AA and BA stacking and the clockwise rotation at the AB stacking. In the main text, we solely report on these average values.

\section{Total energy calculations}
\label{energyAppendixSect}

For all of the observables mentioned in the paper, we use the commensurate cells reported in Table~\ref{Commensurate}. However, when using these same supercells for the total energy calculations from Fig.~\ref{TotalEnergy}, we observed small numerical inaccuracies that lead to erroneous conclusions for the angle-dependence of the energies. We thus calculate the energies using the commensurate cells reported in Table.~\ref{Commensurate_2} that yield smaller variations between lattice mismatches for different twist angles to obtain the energy curves presented in Fig.~\ref{TotalEnergy}. We further adjust the energies using
\begin{equation}
\tilde{E} = E_{\mathrm{atom}} \left[ 1 + \kappa \frac{a_{h-BN}^0}{a_{h-BN}} \right], \qquad \kappa = 0.045
\end{equation} to account for the remaining lattice mismatch noise. We observe that this choice in empirical factor completely smoothens out the noise.

\begin{table}[tbhp]
\resizebox{1.0\columnwidth}{!}{%
\begin{tabular}{|c|c|c|c|c|c|}
\hline
$\theta$ ($^{\circ}$) & ($p$ $q$ $p'$ $q'$) & $a_{h-BN}/a_{h-BN}^0-1$ (\%) & \# atoms & $L_C$ (\AA) & $L_M$ (\AA) \\ \hline
0.1   & 191 136 195 138   & 0.0703  & 329864  & 712.945 & 134.734 \\ \hline
0.25  & 247 175 250 180   & 0.0703  & 549518  & 920.194 & 131.456 \\ \hline
0.303 & 215 217 221 219   & 0.0703  & 570340  & 937.465 & 130.003 \\ \hline
0.560 & 290 234 291 243   & 0.0703  & 842318  & 1139.269 & 119.428 \\ \hline
0.632 & 218 47  220 51    & 0.0703  & 244400  & 613.676 & 115.974 \\ \hline
0.816 & 256 128 265 125   & 0.0703  & 467326  & 848.591 & 106.912 \\ \hline
1.080 & 199 20  200 25    & 0.0703  & 179212  & 525.499 & 94.382 \\ \hline
1.130 & 166 5   167 9     & 0.0703  & 115768  & 422.360 & 92.166 \\ \hline
1.371 & 260 145 257 157   & 0.0703  & 514744  & 890.603 & 82.336 \\ \hline
1.475 & 253 217 268 210   & 0.0703  & 676406  & 1020.921 & 78.532 \\ \hline
1.637 & 185 165 197 159   & 0.0703  & 374776  & 759.931 & 73.124 \\ \hline
1.909 & 196 195 211 187   & 0.0703  & 467216  & 848.491 & 65.269 \\ \hline
2.079 & 195 196 211 187   & 0.0703  & 467216  & 848.491 & 61.076 \\ \hline
3.083 & 238 40  232 57    & 0.0703  & 276122  & 652.287 & 43.680 \\ \hline
3.910 & 178 84  167 103   & 0.0703  & 218782  & 580.622 & 35.141 \\ \hline
5.000 & 123 38  135 24    & 0.0698  & 86576   & 365.246 & 27.850 \\ \hline
10.00 & 74  18  63  35    & 0.0730  & 29062   & 211.620 & 14.171 \\ \hline
14.94 & 62  25  78  2     & 0.0703  & 24526   & 194.403 & 9.520 \\ \hline
20.00 & 31  16  17  31    & 0.0694  & 6980    & 103.708 & 7.140 \\ \hline
25.13 & 29  3   18  18    & 0.0690  & 3818    & 76.701  & 5.701 \\ \hline
29.91 & 30  8   13  27    & 0.0700  & 4906    & 86.946  & 4.808 \\ \hline
\end{tabular}}
\caption{Commensurate angles for G/h-BN and their corresponding four indices in Eq.~(\ref{twistangle}), with more accurate lattice constant $a_{h-BN}$ than Table~\ref{Commensurate}. These systems are used for the total energy calculations reported in Fig.~\ref{TotalEnergy}. $L_C$ corresponds to the length of the commensurate cell, which, for these cases, is larger than $L_M$.}
\label{Commensurate_2}
\end{table}

\end{document}